\documentclass[fleqn,usenatbib]{mnras}

\usepackage{newtxtext,newtxmath}

\usepackage[T1]{fontenc}

\DeclareRobustCommand{\VAN}[3]{#2}
\let\VANthebibliography\thebibliography
\def\thebibliography{\DeclareRobustCommand{\VAN}[3]{##3}\VANthebibliography}

\usepackage{graphicx}	
\usepackage{amsmath}	
\usepackage[dvipsnames]{xcolor}

\newcommand{\hMsun}{h^{-1}\mathrm{M_\odot}}
\newcommand{\hMpc}{h^{-1}\mathrm{Mpc}}
\newcommand{\hGpc}{h^{-1}\mathrm{Gpc}}

\newcommand{\sqdeg}{\mathrm{deg}^2}

\newcommand{\magr}{{}^{0.1}M_r}
\newcommand{\gmr}{{}^{0.1}(g-r)}

\title[Flux-limited Mocks]{Generating mock galaxy catalogues for flux-limited samples like the DESI Bright Galaxy Survey}

\author[A. Smith et al.]{A.~Smith,$^{1}$\thanks{E-mail: alexander.m.smith@durham.ac.uk}
C.~Grove,$^{1}$
S.~Cole,$^{1}$
P.~Norberg,$^{1,2}$
P.~Zarrouk,$^{3}$
S.~Yuan,$^{4}$
J.~Aguilar,$^{5}$
S.~Ahlen,$^{6}$
D.~Brooks,$^{7}$
\newauthor
T.~Claybaugh,$^{5}$
A.~de la Macorra,$^{8}$
P.~Doel,$^{7}$
J.~E.~Forero-Romero,$^{9,10}$
E.~Gazta\~naga,$^{11,12,13}$
\newauthor
S.~Gontcho A Gontcho,$^{5}$
C.~Hahn,$^{14}$
R.~Kehoe,$^{15}$
A.~Kremin,$^{5}$
M.~E.~Levi,$^{5}$
M.~Manera,$^{16,17}$
A.~Meisner,$^{18}$
\newauthor
R.~Miquel,$^{19,17}$
J.~Moustakas,$^{20}$
J.~Nie,$^{21}$
W.~J.~Percival,$^{22,23,24}$
M.~Rezaie,$^{25}$
G.~Rossi,$^{26}$
E.~Sanchez,$^{27}$
\newauthor
H.~Seo,$^{28}$
G.~Tarl\'{e},$^{29}$
and Z.~Zhou$^{21}$
\vspace*{4pt} \\ 
\scriptsize$^{1}$ Institute for Computational Cosmology, Department of Physics, Durham University, South Road, Durham DH1 3LE, UK\vspace*{-2pt}\\
\scriptsize$^{2}$ Centre for Extragalactic Astronomy, Department of Physics, Durham University, South Road, Durham, DH1 3LE, UK\vspace*{-2pt}\\
\scriptsize$^{3}$ Laboratoire de Physique Nucl\'{e}aire et de Hautes Energies (LPNHE), CNRS/IN2P3 \& Sorbonne Universit\'{e}, 4 place Jussieu, 75005 Paris, France\vspace*{-2pt}\\
\scriptsize$^{4}$ SLAC National Accelerator Laboratory, Menlo Park, CA 94305, USA\vspace*{-2pt}\\
\scriptsize$^{5}$ Lawrence Berkeley National Laboratory, 1 Cyclotron Road, Berkeley, CA 94720, USA\vspace*{-2pt}\\
\scriptsize$^{6}$ Physics Dept., Boston University, 590 Commonwealth Avenue, Boston, MA 02215, USA\vspace*{-2pt}\\
\scriptsize$^{7}$ Department of Physics \& Astronomy, University College London, Gower Street, London, WC1E 6BT, UK\vspace*{-2pt}\\
\scriptsize$^{8}$ Instituto de F\'{\i}sica, Universidad Nacional Aut\'{o}noma de M\'{e}xico,  Cd. de M\'{e}xico  C.P. 04510,  M\'{e}xico\vspace*{-2pt}\\
\scriptsize$^{9}$ Departamento de F\'isica, Universidad de los Andes, Cra. 1 No. 18A-10, Edificio Ip, CP 111711, Bogot\'a, Colombia\vspace*{-2pt}\\
\scriptsize$^{10}$ Observatorio Astron\'omico, Universidad de los Andes, Cra. 1 No. 18A-10, Edificio H, CP 111711 Bogot\'a, Colombia\vspace*{-2pt}\\
\scriptsize$^{11}$ Institut d'Estudis Espacials de Catalunya (IEEC), 08034 Barcelona, Spain\vspace*{-2pt}\\
\scriptsize$^{12}$ Institute of Cosmology \& Gravitation, University of Portsmouth, Dennis Sciama Building, Portsmouth, PO1 3FX, UK\vspace*{-2pt}\\
\scriptsize$^{13}$ Institute of Space Sciences, ICE-CSIC, Campus UAB, Carrer de Can Magrans s/n, 08913 Bellaterra, Barcelona, Spain\vspace*{-2pt}\\
\scriptsize$^{14}$ Department of Astrophysical Sciences, Princeton University, Princeton NJ 08544, USA\vspace*{-2pt}\\
\scriptsize$^{15}$ Department of Physics, Southern Methodist University, 3215 Daniel Avenue, Dallas, TX 75275, USA\vspace*{-2pt}\\
\scriptsize$^{16}$ Departament de F\'{i}sica, Serra H\'{u}nter, Universitat Aut\`{o}noma de Barcelona, 08193 Bellaterra (Barcelona), Spain\vspace*{-2pt}\\
\scriptsize$^{17}$ Institut de F\'{i}sica d'Altes Energies (IFAE), The Barcelona Institute of Science and Technology, Campus UAB, 08193 Bellaterra Barcelona, Spain\vspace*{-2pt}\\
\scriptsize$^{18}$ NSF's NOIRLab, 950 N. Cherry Ave., Tucson, AZ 85719, USA\vspace*{-2pt}\\
\scriptsize$^{19}$ Instituci\'{o} Catalana de Recerca i Estudis Avan\c{c}ats, Passeig de Llu\'{\i}s Companys, 23, 08010 Barcelona, Spain\vspace*{-2pt}\\
\scriptsize$^{20}$ Department of Physics and Astronomy, Siena College, 515 Loudon Road, Loudonville, NY 12211, USA\vspace*{-2pt}\\
\scriptsize$^{21}$ National Astronomical Observatories, Chinese Academy of Sciences, A20 Datun Rd., Chaoyang District, Beijing, 100012, P.R. China\vspace*{-2pt}\\
\scriptsize$^{22}$ Department of Physics and Astronomy, University of Waterloo, 200 University Ave W, Waterloo, ON N2L 3G1, Canada\vspace*{-2pt}\\
\scriptsize$^{23}$ Perimeter Institute for Theoretical Physics, 31 Caroline St. North, Waterloo, ON N2L 2Y5, Canada\vspace*{-2pt}\\
\scriptsize$^{24}$ Waterloo Centre for Astrophysics, University of Waterloo, 200 University Ave W, Waterloo, ON N2L 3G1, Canada\vspace*{-2pt}\\
\scriptsize$^{25}$ Department of Physics, Kansas State University, 116 Cardwell Hall, Manhattan, KS 66506, USA\vspace*{-2pt}\\
\scriptsize$^{26}$ Department of Physics and Astronomy, Sejong University, Seoul, 143-747, Korea\vspace*{-2pt}\\
\scriptsize$^{27}$ CIEMAT, Avenida Complutense 40, E-28040 Madrid, Spain\vspace*{-2pt}\\
\scriptsize$^{28}$ Department of Physics \& Astronomy, Ohio University, Athens, OH 45701, USA\vspace*{-2pt}\\
\scriptsize$^{29}$ University of Michigan, Ann Arbor, MI 48109, USA\vspace*{-2pt}\\
}

\date{Accepted XXX. Received YYY; in original form ZZZ}

\pubyear{2024}

\begin{document}
\label{firstpage}
\pagerange{\pageref{firstpage}--\pageref{lastpage}}
\maketitle

\begin{abstract}
Accurate mock galaxy catalogues are crucial to validate analysis pipelines used to constrain dark energy models. We present a fast HOD-fitting method which we apply to the AbacusSummit simulations to create a set of mock catalogues for the DESI Bright Galaxy Survey, which contain $r$-band magnitudes and $(g-r)$ colours. The halo tabulation method fits HODs for different absolute magnitude threshold samples simultaneously, preventing unphysical HOD crossing between samples. We validate the HOD fitting procedure by fitting to real-space clustering measurements and galaxy number densities from the MXXL BGS mock, which was tuned to the SDSS and GAMA surveys. The best-fitting clustering measurements and number densities are mostly within the assumed errors, but the clustering for the faint samples is low on large scales. The best-fitting HOD parameters are robust when fitting to simulations with different realisations of the initial conditions. When varying the cosmology, trends are seen as a function of each cosmological parameter.
We use the best-fitting HOD parameters to create cubic box and cut sky mocks from the AbacusSummit simulations, in a range of cosmologies. As an illustration, we compare the $\magr<-20$ sample of galaxies in the mock with BGS measurements from the DESI one-percent survey. We find good agreement in the number densities, and the projected correlation function is reasonable, with differences that can be improved in the future by fitting directly to BGS clustering measurements. The cubic box and cut-sky mocks in different cosmologies are made publicly available.
\end{abstract}

\begin{keywords}
methods: analytical -- large-scale structure of Universe -- catalogues -- galaxies: statistics
\end{keywords}



\section{Introduction}

The $\Lambda$CDM cosmological model has been very successful at describing the formation and evolution of structure in the Universe \citep{Planck2020}. However, recent tensions have emerged, for example in measurements of the Hubble parameter, between those derived from measurements of the cosmic microwave background and measurements from supernovae in the local Universe \citep[e.g.][]{Verde2019, DiValentino2021, Freedman2021}. In addition,
the nature of dark energy, which drives the accelerated expansion \citep{Riess1998,Perlmutter1999}, and makes up the majority of the energy density of the Universe, remains poorly understood. 

These questions can be probed using the two-point clustering statistics of galaxies in large galaxy surveys. The large-scale structure of the Universe was seeded by primordial fluctuations at early times, which evolved to produce the distribution of galaxies we observe today. Baryon acoustic oscillations (BAO), which propagated through the early Universe, were frozen at the epoch of recombination, leading to a characteristic distance scale in the clustering of galaxies \citep{Cole2005,Eisenstein2005}. This can be used as a standard ruler to measure the expansion history of the Universe. In addition, the peculiar velocity of a galaxy along the line of sight has the effect of shifting the observed redshift of the galaxy \citep{Kaiser1987}. Measuring these redshift space distortions (RSD) in two-point galaxy clustering statistics provides a test of general relativity, and constrains modified gravity models \citep{Guzzo2008}. 

The Dark Energy Spectroscopic Instrument \citep[DESI;][]{DESI2016science,DESI2016instrument,DESI2022instrument} is currently undertaking a 5-year survey that will measure the spectra of approximately 40 million galaxies and quasars between $0<z<3.5$, and which aims to place our best constraints on models of dark energy. DESI is targeting several galaxy tracers over this redshift range, including luminous red galaxies (LRGs), emission line galaxies (ELGs) and quasars (QSOs). During bright time, DESI is conducting the Bright Galaxy Survey (BGS), in addition to the Milky Way Survey (MWS) of stars within the Milky Way galaxy. The BGS is a flux-limited survey of low redshift galaxies ($z \sim 0.2$). The BGS-BRIGHT sample will contain over 10 million galaxies brighter than $r=19.5$. The secondary BGS-FAINT sample will extend this magnitude limit to $r=20.175$, with additional cuts based on fibre magnitude and colour to ensure a high redshift success rate \citep{Hahn2023}. The BGS will provide a highly complete galaxy catalogue not only for cosmological measurements, but also for studies of galaxy formation and evolution. The first DESI data from the survey validation program \citep{DESI2023SurveyValidation} has recently been made publicly available \citep{DESI2023DataRelease}.

The use of accurate simulated mock galaxy catalogues are critical for large surveys like DESI, to aid in survey design and assess survey strategies \citep{Looser2021}. In addition, mocks are needed to test and optimise the key data analysis pipelines that are designed to handle the large volumes of data. Synthetic datasets can be used to test the recovery of key statistics from survey data, investigating potential systematic errors and ensuring the unbiased measurement of cosmological parameters. Since the volume probed by modern galaxy surveys is so large, these simulations also need to cover extremely large cosmological volumes. For covariance matrices, approximate or low resolution simulations are used in order to generate many thousands of mocks.

To create realistic mock galaxy catalogues, it is necessary to accurately model the link between galaxies and their host dark matter halos. The galaxy-halo connection has been a subject of research for several decades \citep{Cole1989,Mo_White_1996,Cooray2002,Wechsler2018}, with many formulations from simple Halo Occupation Distributions (HODs) to more complex treatments that aim to model the impact of various physical processes on galaxy formation in hydrodynamic and semi-analytic simulations.

Hydrodynamic galaxy formation simulations model the galaxy-halo connection directly by including both dark-matter and baryonic components interacting over time directly through gravity and other forces \citep[e.g.][]{Vogelsberger2014,Crain2015,McCarthy2017,Lee2021}.
However, these simulations are very computationally expensive, limiting the volume that can be simulated in a reasonable amount of time.
While it is recently becoming possible to run hydrodynamical simulations with Gpc box sizes \citep[e.g. the FLAMINGO simulations,][]{Schye2023} for the simulation volumes required for large surveys like DESI, it is typically only feasible to run dark-matter-only simulations, using methods to paint galaxies onto the dark matter halos. 

Semi-analytic galaxy formation models can model the formation and evolution of galaxies in an existing dark-matter-only $N$-body simulation. Physically-informed models simulate a variety of processes such as star formation, feedback, and radiative heating and cooling \citep[e.g.][]{Croton2006,Lacey2016,Henriques2020}. 
There are a large number of degrees of freedom and assumptions underlying the physical models which are constrained to match observations. 
While this can be applied to a large dark-matter-only simulation, high resolution halo merger trees are required. 

Sub-halo abundance matching (SHAM) techniques can be used to link galaxies to sub-halos, based on a ranking of sub-halo and galaxy properties \citep[e.g.][]{Conroy2006,Reddick2013,Prada2023}. 
To reproduce galaxy clustering statistics, the relationship between the sub-halo and galaxy properties must include some scatter \citep{Tasitsiomi2004}. SHAM techniques require high simulation resolution in order to resolve sub-halos sufficiently, otherwise halos may be over-merged and systematic errors can be produced in the clustering \citep{Guo2014}. The SHAM algorithm has been extended to e.g. include the satellite fraction \citep{Favole2016}, assembly bias \citep{Contreras2021} and observational systematics \citep{Yu2022}.

Another method is the conditional luminosity function (CLF), which was introduced by \citet{Yang2003}. The CLF describes the halo occupation statistics in terms of galaxy luminosity, by modelling the luminosity function of galaxies residing in halos of mass $M$. The CLF has previously been applied to SDSS data in order to obtain cosmological constraints \citep{vandenBosch2013,Cacciato2013,More2013}.

The HOD is a more empirical method for modelling the galaxy-halo connection. Instead of assuming that galaxies are formed by specific physical processes which affect their abundance and properties, HOD methods simply assume that the abundance of galaxies is informed by the host halo properties without reference to a physical model. This abstracts away much of the underlying physics but is sufficient if one only cares about knowing what the galaxy-halo connection is, rather than why it takes that form \citep{Berlind2002}. The HOD technique was applied to SDSS data in \citet{Zehavi2011}.

It is relatively simple to use a HOD model for a single galaxy sample to add galaxies to the dark matter halos of a simulation. However, for the BGS, we also want to assign galaxy properties, like magnitudes.
This can be done using a set of `nested' HODs, as has been done previously in
\citet{Skibba2006,Smith2017,Paul2019,Smith2022MXXL}.
In \citet{Smith2017,Smith2022MXXL} a method based on \citet{Skibba2006} was developed to assign each mock galaxy a SDSS $r$-band magnitude using a set of `nested' HODs for different magnitude thresholds. However, these SDSS HODs were measured for each magnitude-threshold sample independently, and needed to be modified to prevent
the HODs from crossing over each other. If one galaxy sample is a subset of another, it is unphysical for the two HOD curves to cross. The average number of galaxies in halos of mass $M$ that are brighter than a certain magnitude threshold must always be a monotonic function of magnitude. In \citet{Paul2019}, nested HOD for different magnitude thresholds were fitted to SDSS clustering measurements, which were
used to construct SDSS mocks with multi-band luminosities \citep{Paranjape2021}.

In the analysis of the DESI BGS data, a set of accurate mock galaxy catalogues are required that reproduce the BGS luminosity function and magnitude-dependent clustering. The mocks we produce here are a first step towards achieving this aim, by developing a mock creation pipeline based on existing datasets, which can then be extended to the real DESI data as it becomes available. In addition to validating the models used in the standard two-point clustering analyses, these mocks can be used for assessing new statistics that can be measured in the BGS, e.g. multi-tracer and three-point statistics. The mocks we make are created from AbacusSummit, which is a set of simulations that includes boxes run in a range of different cosmologies. This allows us to study how the galaxy halo connection varies with cosmology, in addition to testing the impact on the cosmological analyses. These mocks are made publicly available, and have been used as part of the DESI year-1 BAO \citep{DESI2024BAO} and RSD analyses \citep{DESI2024RSD}.

To create the mocks, we simultaneously fit HODs to galaxy clustering and number density constraints from multiple magnitude-threshold samples, preventing any unphysical crossing of the HODs between samples. We fit to measurements from the previous MXXL BGS mock catalogue, which provides an accurate estimate of the expected clustering of the BGS survey. The volume of the 1\% survey is too small and affected by cosmic variance for HOD fitting, but in future work we will apply this method to the larger DESI year-1 dataset. Our goal is to measure a set of HODs which can be used to create mocks that reproduce the clustering and number density of the BGS.
We describe the HOD-fitting methodology, and show the results from fitting to MXXL clustering measurements. 
We also discuss the limitations of the method, and the modifications and improvements that are needed in future work to apply the method directly to BGS data.
The outline of this paper is as follows. In Section~\ref{sec:hod} we describe our HOD model for linking BGS galaxies to dark matter halos. The tabulation method for fast HOD fitting is described in Section~\ref{sec:tabulation_method}.
In Section~\ref{sec:hod_fitting_section}, we fit HODs to the AbacusSummit simulations, testing the robustness of the method and explore how the HOD parameters vary for simulations with different cosmologies.
The method for creating cubic box and cut-sky mocks is described in Section~\ref{sec:mock_creation}.
Finally, we summarise our conclusions in Section~\ref{sec:conclusions}.

\section{Linking galaxies and dark matter}
\label{sec:hod}

\subsection{Halo Occupation Distribution}

Galaxies are biased tracers of the underlying matter density field of the Universe. In the halo model, it is assumed that all galaxies reside within larger dark matter halos. The link between galaxies and halos can be modelled using a halo occupation distribution, which specifies the average number of galaxies within each halo. In its simplest form, the HOD is purely a function of the halo mass, and the form of the HOD is often split up into two components, to model the abundance of central and satellite galaxies. The specific choice of form for the HOD may depend on the selection of the galaxy sample being modelled. For samples such as luminous red galaxies (LRGs), where there is a correlation (with scatter) between halo mass and stellar mass (or galaxy luminosity), the HOD of central galaxies is modelled with a smoothed step function that approaches one at high masses \citep{Zheng2005}. For emission line galaxies (ELGs), which are star forming, the central HOD is instead modelled with a Gaussian-like distribution for the central galaxy occupation, decreasing at high masses since these halos often host red elliptical central galaxies, with low star formation rates \citep[e.g.][]{Avila2020}. The occupation number of satellite galaxies is often modelled as a power law, with a cut-off at low masses.

More complex versions of HOD methods exist to account for the relationship between other halo properties and galaxy abundance, known as assembly bias. Decorated HODs have been explored as methods to improve HODs, which add dependence of the HOD on several more parameters such as formation time and concentration \citep{Paranjape2015,Hearin2016,Yuan2022AbacusHOD}. 
For example, \citep{Alam2024} found some evidence of assembly bias in the GAMA data, using an extended HOD model developed in \citet{Alam2021}.

The central galaxy is usually assumed to be at the centre of the halo, and there are different methods of modelling the placement of satellite galaxies. The satellites are often assumed to follow a Navarro-Frenk-White (NFW) profile \citep{Navarro1996,Navarro1997}. Alternatively, in a N-body simulation, the dark matter particles within each halo can be used as tracers for the satellites.

\subsection{HOD parametrisation}
\label{sec:hod_parametrisation}

In this work, we use a HOD model of the same form as \citet{Smith2017} to model the HOD of DESI BGS galaxies, where there is a correlation between galaxy luminosity and halo mass. This is closely related to \citet{Zheng2005}, and is described by five parameters in total: two for the central galaxy component and three for the satellites. The occupation number for central galaxies brighter than luminosity $L$ is modelled using a smoothed step function as a function of halo mass, $M$, 
\begin{equation}
\label{eq:hod_cen}
    \left\langle N_{\mathrm{cen}}(>L \mid M)\right\rangle=\frac{1}{2}\left[1+F\left(\frac{\log M-\log M_{\min }(L)}{\sigma_{\log M}(L)}\right)\right] ,
\end{equation}
where $M_{\textrm{min}}$ is the position of the step and $\sigma_{\log M}$ the width. 

In the standard HOD formalism of \citet{Zheng2005}, the shape of the smooth step function is modelled as an error function. We replace this with an equivalent function, $F(x)$, which is defined as 
$F(x)=2 \int_{0}^{x} S\left(x^{\prime}\right) d x^{\prime}$, where $S(x^{\prime})$ is a pseudo-Gaussian function spline kernel function,
\begin{align}
\label{eq:spline}
\mathrm{spline}(x)         & = 1 - 6\mathopen|x\mathclose|^{2} + 6 \mathopen|x\mathclose|^{3}           & \mathopen|x\mathclose| \leq 0.5 \nonumber\\
& = 2(1-\mathopen|x\mathclose|)^{3} & 0.5 < \mathopen|x\mathclose| \leq 1\\
& = 0 & 1< \mathopen|x\mathclose|, \nonumber
\end{align}
which has been normalised and rescaled to mean $\mu=0$ and variance $\sigma^2=1/2$. This is done following
\begin{equation}
S(x,\mu,\sigma) = \frac{4/3}{\sigma\sqrt{12}} \mathrm{spline} \left( \frac{x - \mu}{\sigma \sqrt{12}} \right) .
\end{equation}

This has the advantage that the tails of the pseudo-Gaussian function are truncated to exactly zero, helping to prevent unphysical crossing of the HODs of different absolute magnitude threshold samples.

The satellite HOD is a power law weighted by the central HOD,
\begin{equation}
\label{eq:hod_sat}
\left\langle N_{\mathrm{sat}}(>L \mid M)\right\rangle=\left\langle N_{\mathrm{cen}}(>L \mid M)\right\rangle\left(\frac{M-M_{0}(L)}{M_{1}^{\prime}(L)}\right)^{\alpha(L)} ,
\end{equation}
where $M_0$ is the low mass cutoff mass scale, $M_1^{\prime}$ the normalisation, and $\alpha$ the power law slope.

\subsection{Varying HOD parameters with magnitude}
\label{sec:hod_magnitude_parametrisation}

A central assumption in this implementation is that HOD parameters are defined to smoothly vary with an absolute magnitude limit. This is a reasonable assumption because we are not modelling a galaxy population that is sensitive to an on/off state such as star formation. This allows us to define the HOD at any magnitude and therefore to populate the mock galaxy catalogue with galaxies that have realistic luminosities, rather than populating a mock that represents a fixed magnitude cut. Fitting HOD parameters to smoothly varying curves presents some new challenges and there are several ways in which this could be implemented.

One option is to first independently fit HOD parameters for a set of different magnitude threshold samples. Smooth curves can then be fit to each parameter, as a function of magnitude. This method has the benefit of being relatively quick and simple, as one only has to repeat the standard HOD fitting procedure several times, once for each magnitude limit. However, there is no guarantee that the best-fitting HOD parameter values can be well approximated by a smooth curve as a function of magnitude. The HOD parameters produced by these smooth functions also may not preserve the target galaxy clustering which has also been fitted. This is the approach that was taken in \citet{Smith2017}, but adjustments were required to prevent unphysical crossing between HODs of bright and faint magnitude threshold samples.

The fitting method chosen in this paper is to first parametrise how each HOD parameter varies with magnitude before fitting. The meta-parameters describing these functions are then fit, meaning that no further adjustment needs to be made to the fitted HOD parameters. Each HOD parameter, and therefore the shape of the total HOD, can be obtained for any magnitude limit, using this parametrisation. One potential pitfall of this method is that if the parameterised curves are too constrained, then a good fit to the target clustering may not be found and therefore different forms for the curves may need to be tested in order to produce robust results. Another issue is that fitting all of the meta-parameters at once is a higher dimensional fit than fitting parameters for individual magnitudes separately. This takes up more computing time to complete the fit and the large number of parameters means that the fitting procedure is more likely to end up in a local minimum than be able to locate the global minimum. We attempted to overcome this by running the fitting procedure several times from different starting points in the parameter space. This provides more confidence that the best fit HOD parameters are stable without needing to allocate vastly greater computational resources to the problem. 

\begin{figure}
    \centering
    \includegraphics[width=\columnwidth]{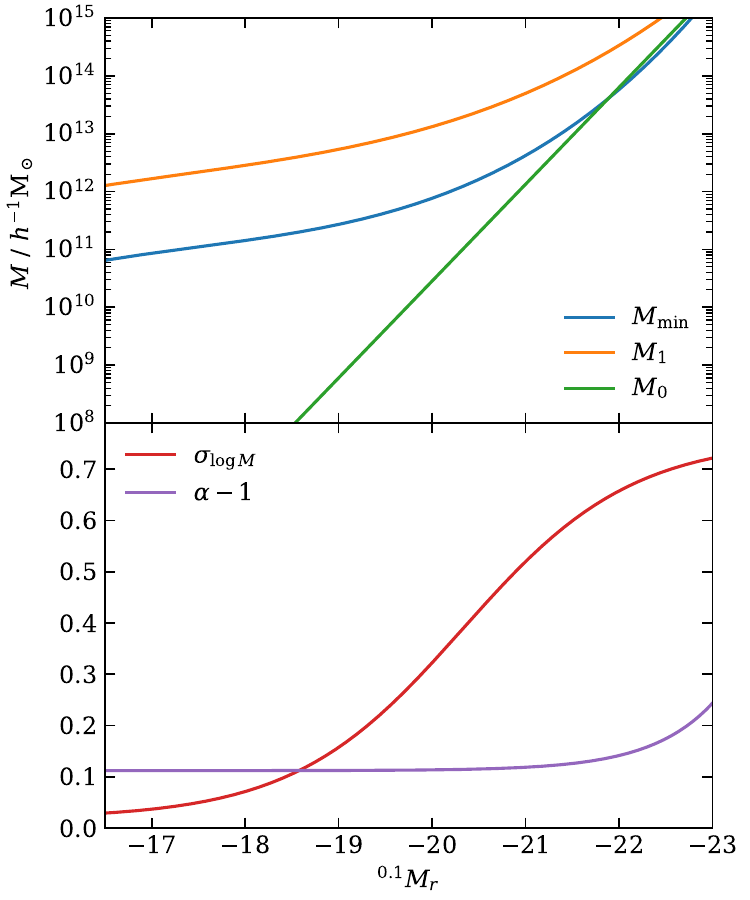}
    \caption{The best-fitting HOD parameters, as a function of magnitude, for the AbacusSummit Planck `c000' cosmology box with initial conditions `ph000', illustrating the functional forms of Eqs.~\ref{eq:hod_Mmin}-\ref{eq:hod_alpha}. The upper panel shows the mass parameters $M_{\textrm{min}}$, $M_0$, and $M_1$ on a logarithmic scale, in blue, orange and green, respectively. The lower panel shows the parameters $\sigma_{\textrm{log}M}$ in red and $\alpha$ in purple. We plot $\alpha-1$ so that both parameters fall within the same y-axis range.
    }
    \label{fig:hod_fit}
\end{figure}

The parametrised forms of the 5 HOD parameters are shown below, where each HOD parameter is written as a function of absolute magnitude. In total, there are 17 parameters. The functional forms were chosen as they approximately match the measured HOD parameters when HOD fits are done independently for each magnitude threshold. $\log M_0$ is described by a linear function of magnitude, $\sigma_{\log M}$ is described by a smoothly varying step function and $\alpha$ is described by a constant value with an exponential component at bright magnitudes. We describe the mass parameters $\log M_\mathrm{min}$ and $\log M_1$ with cubic functions. This is a different functional form then used by e.g. \citet{Zehavi2011}, but we find that it produces better fits. The five functions are given by
\begin{align}
\label{eq:hod_Mmin}
\textrm{log} M_{\textrm{min}} &= 12 + A_\mathrm{min} + B_\mathrm{min} M_r' +  C_\mathrm{min} M_r'^2 +  D_\mathrm{min} M_r'^3 \\[1ex]
\sigma_{\textrm{log}M} &= A_\sigma + \frac{B_\sigma - A_\sigma}{1+\exp({C_\sigma(M_r'+D_\sigma))}} \label{eq:hod_sigma} \\[1ex]
\textrm{log} M_0 &= 11 + A_0 + B_0 M_r' \label{eq:hod_M0} \\[1ex]
\textrm{log} M_1 &= 12 + A_1 + B_1 M_r' +  C_1 M_r'^2 +  D_1 M_r'^3 \label{eq:hod_M1} \\[1ex]
\alpha &= A_\alpha + B_\alpha^{-M_r' + C_\alpha}, \label{eq:hod_alpha}
\end{align}
For conciseness we define $M_r' = \magr + 20$, where $\magr$ is the $r$-band absolute magnitude, $k$-corrected to a reference redshift $z=0.1$ (see Section~\ref{sec:cut_sky_mocks}). 
The HOD parameters as a function of magnitude are illustrated in Figure~\ref{fig:hod_fit}.
These 5 functions are similar to how the SDSS HODs were parametrised in order to construct the MXXL mock catalogue of \citet{Smith2017}. A different functional form was used for $M_\mathrm{min}$ and $M_1$, which only had 3 free parameters \citep{Zehavi2011}. However, in this work we find that cubic functions with an extra free parameter are able to produce better HOD fits.

\subsection{Preventing HOD crossing}

For two galaxy samples with lower luminosity limits/thresholds $L_1$ and $L_2$ (with $L_2 > L_1$), it must always be true that the number of galaxies $N_\mathrm{gal}(>L_1) \ge N_\mathrm{gal}(>L_2)$, since the galaxies with luminosity $L>L_2$ are a subset of those with $L>L_1$. 
It is therefore unphysical for the HODs to cross, as this condition would no longer be true, and would require a negative number of galaxies in the range $L_1<L<L_2$.
As already discussed, this motivates the choice of HOD parametrisation, which uses a pseudo-Gaussian function to truncate the tail of the central HODs at low masses. 

This consideration can also be built into the HOD fitting procedure by adding a high penalty to the likelihood if the HODs cross.

An alternative approach was adopted in \citep{Paul2019}, where HODs were fit to magnitude-threshold samples from the SDSS data. Here, rather than fitting directly to the clustering of magnitude-threshold samples, the fits were done to clustering measurements in magnitude bins. A joint likelihood was constructed which enforces the monotonicity of the HODs statistically.

\section{Halo tabulation}
\label{sec:tabulation_method}

In order to efficiently explore the parameter space of possible HOD forms, a fast method to evaluate the expected galaxy clustering is needed. We have implemented a version of the halo paircounting method introduced by \citet{Neistein2011} and \citet{Zheng2016}. This method tabulates the dark matter halo paircounts by mass and separation, then the average effect of using a particular HOD can be estimated by reweighting elements of these tables and summing the contributions to the clustering. 
The \citet{Zheng2016} tabulation method was also used in the HOD fitting of \citet{Paul2019}.

Using this method means that the computationally expensive paircounting routine needs only to be run once in total, instead of once for each set of HOD parameters. The paircounting is run on the halos instead of the galaxies, and the larger number of halos means that there is a greater fixed cost to using this method. However when many evaluations are needed in the fitting procedure, the much quicker time for each evaluation leads to a quicker runtime in total.

Alternatives to explicitly calculating the clustering produced by a set of HOD parameters can include using emulators or analytic methods \citep{Kwan2015,Zhai2019,Yuan2022}, but these are not guaranteed to produce accurate results for a given simulation.

This tabulation method assumes that occupancy of a halo depends on only one variable, the halo mass. We can bin halo paircounts by different combinations of halo mass and then reweight these paircounts to account for changes in the HOD. It is possible to include other halo parameters by mixing them with the halo mass to create an {\emph{effective mass}} which maintains the HOD function as dependent on a single variable \citep{Yuan2018}, however this method is not tested here. HODs with two or more input variables can be used, but this increases the dimension of the tabulated paircounts and leads to an exponential increase in code runtime.

\subsection{Tabulation Method}

Here we give an overview of the tabulation method, but a more detailed description can be found in \citet{CameronThesis}, which includes additional run time and accuracy tests. 

We aim to estimate the real-space two-point correlation function, $\xi(r)$, which can be written as
\begin{equation}
\label{eq:2pcf}
\xi(r) = \frac{GG(r)}{RR(r)} -  1,
\end{equation}
where $GG(r)$ and $RR(r)$ are the normalized galaxy-galaxy and random-random pair counts respectively, in bins of separation $r$. For our periodic simulation boxes, $RR(r)$ can be calculated analytically. 
We work in real space since this simplifies the halo tabulation.

The galaxy pair counts can be split into components from central and satellite galaxies,
\begin{equation} 
\label{eq:gg_pair_count_sum}
GG(r) =  CC(r) + CS(r) + SS_\mathrm{2halo}(r) + SS_\mathrm{1halo}(r),
\end{equation}
where $CC$ and $CS$ represent the central-central, central-satellite galaxy pair counts, respectively. The satellite-satellite pair counts are further split into the 2-halo term, where the satellite occupy different halos, $SS_\mathrm{2halo}$, and a 1-halo term of satellites that reside in the same halo, $SS_\mathrm{2halo}$.

Each of these terms can be expressed as a sum over galaxy pairs which reside in halos of different masses. For example, for the central-central pairs,
\begin{equation}
CC(r) = \sum_{M_{ij}}{CC(M_{ij},r)},
\end{equation}
with similar expressions for the other terms in Eq.~\ref{eq:gg_pair_count_sum}. Here, $CC(M_{ij},r)$ is the number of central-central pairs which live in halos of masses $M_i$ and $M_j$ (which we write as $M_{ij}$). We use evenly spaced logarithmic mass bins, and pairs in the sum must not be double counted. Each of these sums can be evaluated from the halo pair counts and the HOD model.

The HOD model we use, which is described in Section~\ref{sec:hod_parametrisation}~and~\ref{sec:hod_magnitude_parametrisation}, provides the mean number of central galaxies in halos of mass $M$, $N_\mathrm{cen}(M)$ (where $0 \le N_\mathrm{cen}(M) \le 1$), and the mean number of satellite galaxies, $N_\mathrm{sat}(M)$.
We assume that there does not have to be a central galaxy in a halo in order for it to host satellite galaxies. This assumption simplifies the tabulation calculations significantly as otherwise one would have to consider a larger number of correlations from halos explicitly with and without central galaxies. 
In practice the majority of satellite galaxies are placed in halos containing a central galaxy as they are preferentially placed in high-mass halos. 
The low mass end of the HOD is dominated by the central term, making it very unlikely to have satellites without a central.
Central galaxies are positioned at the centre of the halo and satellite galaxies are positioned randomly according to an NFW profile around the halo. The number of satellite galaxies per halo is chosen according to a Poisson distribution with mean $N_{\rm sat}(M)$. 

The central galaxies will share positions with the halos as they are placed at the centre, therefore the distribution of potential central galaxy positions is sampled by the halo centres themselves. For the satellite galaxies we place a fixed number of tracer particles around each halo to represent where satellite galaxies would be placed. In this work, we use 3 satellite tracers per halo.

We represent the number of pairs of halos in mass bins $M_i$ and $M_j$ as $W^{\rm CC}_{ij}(r)$. Using the satellite tracers for each halo, we similarly define $W^{\rm CS}_{ij}(r)$, $W^{\rm SS1}_{ij}(r)$ and $W^{\rm SS2}_{ij}(r)$, which are the halo-satellite pair counts, and the satellite-satellite counts divided into 1-halo and 2-halo terms.

The galaxy central-central pair counts can then be evaluated from the sum
\begin{equation}
 CC(r) = \sum_{M_{ij}} N_{\rm cen}(M_i) N_{\rm cen}(M_j) W^{\rm CC}_{ij}(r).
\end{equation}
Similar expressions can be written for the central-satellite and satellite-satellite 2-halo terms, 
\begin{equation}
\label{eq:halo_cs}
 CS(r) = \frac{1}{T} \sum_{M_{ij}} N_{\rm cen}(M_i) N_{\rm sat}(M_j) W^{\rm CS}_{ij}(r) 
\end{equation}
\begin{equation}
\label{eq:halo_ss2}
 SS_\mathrm{2halo}(r) = \frac{1}{T^2} \sum_{M_{ij}} N_{\rm sat}(M_i) N_{\rm sat}(M_j) W^{\rm SS2}_{ij}(r),
\end{equation}
where $T$ is the number of satellite tracers per halo. Note the different dependence on the number of satellite tracer particles between Equation~\ref{eq:halo_cs} and \ref{eq:halo_ss2}. 

In practice, $T=1$ is sufficient for accurate calculation of both the central-satellite term and the satellite-satellite two-halo term. For the one-halo term, we must use more than one tracer particle per halo to sample the possible galaxy pairs. The number of one-halo satellite galaxy pairs sampled per halo when using $T$ tracer particles per halo is $T(T-1)/2$.
Therefore in this case we relate the galaxy and halo pair counts using
\begin{equation}
SS_\mathrm{1halo}(r) = \frac{1}{T(T-1)} \sum_{M_{ij}} N_{\rm sat}(M_i) N_{\rm sat}(M_j) W^{\rm SS1}_{ij}(r).
\end{equation}
Here all the off-diagonal terms are zero as there are no pairs of satellites between halos of different masses in the one-halo term. In our calculations, we use $T=3$ satellite tracers per halo.
 
 In each case the halo paircounts are static and only depend on the halo catalogue, meanwhile the HOD factors can vary allowing us to fit HOD parameters rapidly by evaluating the sums shown above.

These estimated galaxy paircounts can be combined with analytic randoms to produce the expected correlation function as shown in Eq.~\ref{eq:2pcf}.
The analytic expression for the number of random pairs is
\begin{equation}
    RR(r) = \frac{\textrm{d}V(r)}{V}N(N-1), 
\end{equation}
where $\textrm{d}V(r)$ is the volume of a spherical shell with minimum and maximum radii corresponding to the edges of the radial bin in the clustering calculation, $V$ is the total volume of the simulation box, and $N$ is the expected number of galaxies in the whole simulation volume which can be estimated by applying the HOD to the halo mass function. The number of pairs of randoms vary as the HOD parameters change because the expected total number of galaxies that will be populated is altered. See section~4.5.3 of \citet{CameronThesis} for tests of the tabulation accuracy.

\subsection{Halo Subsampling}

\begin{figure}
    \includegraphics[width=\columnwidth]{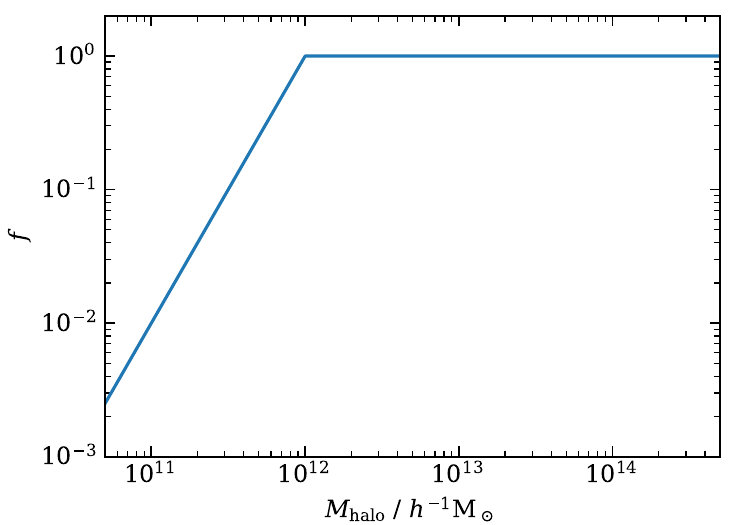}
    \caption{Fraction of halos that are subsampled, $f$, as a function of halo mass. Halos are subsamples in order to speed up the computation of the halo pair counts. Subsampling is only done for halos with mass $M < 10^{12}~\hMsun$; all halos are used above this. At $M=10^{11}~\hMsun$, the
    subsampled fraction is 1\%. }
    \label{fig:subsampling_fraction}
\end{figure}

The halos used in the paircounting were subsampled in order to speed up computation of the tables of the binned paircounts. There are vastly more low mass halos than high mass halos in the halo catalogue and low mass halos are proportionally occupied by fewer galaxies. This means that HOD fits can be robust to a subsampling of low mass halos. By testing of different subsampling forms and mass cutoffs, it was found that the runtime could be improved by a factor of ten with negligible impacts to the measured number density and clustering. The form of the subsampling was a function of halo mass, with no subsampling at masses above $10^{12} h^{-1} \textrm{M}_\odot $. The subsample fraction $f$ as a function of halo mass $M$ is set as 
$\log_{10}f = \mathrm{min} \{0, 2(\log_{10}M - 12) \}$, with $M$ in units $\hMsun$. 
This function is shown in Figure~\ref{fig:subsampling_fraction}.
We checked with one of the simulations that changes to the number densities and clustering of HOD fits introduced by the subsampling were smaller than 0.1\%. This is at a level much smaller than the precision we aim to achieve in the HOD fits, so halo subsampling will not bias our results.

\subsection{Unresolved Halos}
\label{sec:unresolved}

\begin{figure}
    \centering
    \includegraphics[width=\columnwidth]{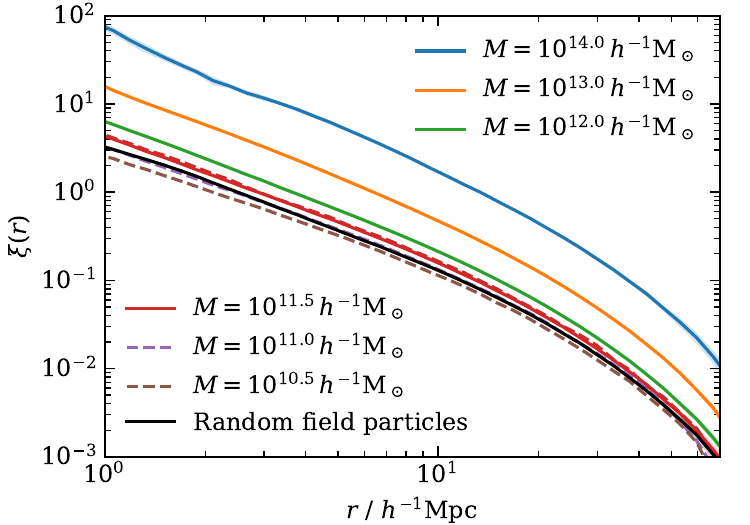}
    \caption{The real-space 2-point clustering of field particles along with halos of different masses from the base AbacusSummit simulation. Correlation functions are truncated at small scales where there are insufficient pairs for it to be measured. The solid lines show the average measurements from the 25 `base' resolution boxes with side length $2~\hGpc$, and the shaded region indicates the $1\sigma$ scatter. Dashed lines show measurements from one AbacusSummit `high' simulation, which is a $1~\hGpc$ box in the same cosmology, but with $\sim 6$ times better mass resolution.}
    \label{fig:low_mass_clustering}
\end{figure}

A halo mass cut at $10^{11}~\hMsun$ was applied to the AbacusSummit halo catalogues we use in this work (see Section~\ref{sec:abacussummit_simulations}) because halos are not sufficiently resolved below this limit. Randomly selected field particles were used as the locations of halos below this mass cut and these were not given satellite tracers because the satellite contribution from the HOD used in this work is negligible at low masses. 
Figure~\ref{fig:low_mass_clustering} shows the clustering of the field particles, compared to the resolved halos of the simulation. The field particles have a similar clustering amplitude as the $10^{11}~\hMpc$ halos, showing that using the field particles is a reasonable approximation for extending the HOD below the mass resolution limit of the halos. While the clustering amplitude decreases further for halos less massive than $10^{11}~\hMpc$, only a small fraction of BGS galaxies reside in such low mass halos.

\section{HOD fitting procedure}
\label{sec:hod_fitting_section}

In this section we describe the HOD fitting procedure, which we apply to the AbacusSummit simulations. The simulations and data used in this work are described in Section~\ref{sec:simulations_data}, and the HOD fitting procedure in Section~\ref{sec:hod_fitting}. The best-fitting HODs in Planck cosmology are discussed in Section~\ref{sec:best_fitting_hods}, and the dependence of these results on cosmology in Section~\ref{sec:varying_cosmology}.

\subsection{Simulations and data}
\label{sec:simulations_data}

The simulations and data used in this work are described below. We fit HODs using the AbacusSummit simulations (Section~\ref{sec:abacussummit_simulations}) to clustering measurements obtained from the MXXL mock catalogue (Section~\ref{sec:mxxl_mock}). The mocks constructed from AbacusSummit are then compared to the DESI BGS One-percent survey data (Section~\ref{sec:desi_bgs_data}). The HODs are fit to MXXL, which in turn was fit to SDSS and GAMA data \citep{Smith2017}, rather than the BGS data directly because the one-percent survey is somewhat smaller, shallower and less complete than GAMA. Direct comparisons between the mocks and data are enabled by using the same magnitude selection, using the same $k$- and $E$-corrections.

\subsubsection{AbacusSummit simulations}
\label{sec:abacussummit_simulations}
AbacusSummit\footnote{\href{https://abacussummit.readthedocs.io/}{https://abacussummit.readthedocs.io/}}\citep{AbacusSummit} is a suite of $N$-body simulations run using the \textsc{abacus} code \citep{Abacus_Code} 
on the Summit supercomputer. There are 150 simulations that have been run with different cosmologies, resolutions and initial conditions. In this work we use the `base' resolution simulations, which contain $6912^3$ particles in a cubic box of side length $2~\hGpc$, with a particle mass of $2.11 \times 10^{9}~\hMsun$.

The AbacusSummit simulations were run from 2LPT initial conditions at $z=99$. The primary `c000' cosmology corresponds to Planck 2018 $\Lambda$CDM results \citep{Planck2020}, with cosmological parameters $h = 0.6736$, $\Omega_\mathrm{cdm} h^2 = 0.1200$,  $\Omega_\mathrm{b} h^2 = 0.02237$, $\sigma_8 = 0.8114$ and $n_{s} = 0.9649$. A total of 25 boxes were produced in the primary cosmology, with different initial conditions. The secondary cosmologies are selected to match existing flagship $N$-body simulation projects, in addition to a wide emulator grid around the primary cosmology.
In the emulator grid, the parameters $\sigma_8$, $\Omega_\mathrm{cdm}h^2$, $n_s$, $\Omega_\mathrm{b}h^2$, $w_0$, $w_a$, $N_{ur}$ and $\alpha_s$ are all varied around the Planck values, where $\alpha_s$ is a running spectral index, $N_{ur} \sim 2$ is the effective number of massless neutrinos, and $w_0$ and $w_a$ set an evolving dark energy equation of state of the form $w(z)=w_0 + w_a(1-a)$. These cosmologies include a single species of massive 60~meV neutrinos, with $\Omega_\mathrm{ncdm}h^2=6.4420 \times 10^{-4}$. 
In the primary Planck cosmology, $\alpha_s=0$, $N_{ur}=2.0328$, $w_0=-1$ and $w_a=0$ \citep{AbacusSummit}.

Halo catalogues are the primary data product from AbacusSummit, with the full set of all particles being too large to store efficiently. Halo finding in AbacusSummit was performed on the fly using CompaSO \citep{Hadzhiyska2022CompaSO}, a spherical overdensity method. A friends-of-friends \citep{Davis1985} algorithm is first applied to identify `L0' halos. `L1' halos are found using a a spherical overdensity algorithm, which finds particles within a density threshold $\Delta=200$, while `L2' subhalos are found using $\Delta=800$. For each halo, we use use a halo mass which is defined as the total number of particles within the L1 halo (corresponding to $M_\mathrm{200m}$). The L2 cores of each halo are used to find the positions and velocities. The CompaSO algorithm can deblend halos, producing unphysical objects. We use the cleaned CompaSO halo catalogues which are processed using merger tree information to fix these issues \citep[see][]{Hadzhiyska2022CompaSO}.

Subsets of the particles are available at certain snapshots. This is split into an `A' sample (3\% of the particles) and a `B' sample (7\% of the particles). We use the random subset of field particles (which do not exist in halos) to extend the halo catalogue below the mass resolution of the simulation (see Section~\ref{sec:unresolved}).

\subsubsection{MXXL BGS mock}
\label{sec:mxxl_mock}
The Millennium-XXL (MXXL) simulation \citep{Angulo2012} is a dark-matter-only simulation that is a successor to the Millennium simulation \citep{Springel2005}, with a much larger box size of $3~\hGpc$. The particle mass is $6.17\times10^9~\hMpc$, and the simulation was run in the same WMAP1 cosmology as the original Millennium simulation, with $\Omega_\mathrm{m}=0.25$, $\Omega_\mathrm{b}=0.045$
$\Omega_\Lambda=0.75$, $\sigma_8=0.9$, $h=0.73$ and $n_s=1$ \citep{Spergel2003}. At each simulation snapshot, dark matter halos are first identified using a friends-of-friends algorithm, with bound substructures identified using \textsc{subfind}. To construct a halo merger tree, the 15 most bound particles of each halo were found, and the descendent is the halo at the next snapshot containing the majority of these particles.

The MXXL simulation was previously used to create a mock galaxy catalogue for the BGS, as described in \citet{Smith2017,Smith2022MXXL}. Halos were first interpolated between simulation snapshots to build a halo lightcone, and were then populated with galaxies using a set of `nested' HODs for different $r$-band magnitude thresholds, which were measured from the SDSS survey. The HODs were evolved with redshift to reproduce an evolving $r$-band target luminosity function from SDSS and GAMA. A Monte Carlo method is used to assign each galaxy a luminosity from the nested HODs, and satellites are positioned following a NFW profile. The galaxies are also randomly assigned $g-r$ colours from a parametrisation of the GAMA colour-magnitude diagram. In addition to creating a lightcone mock, the same HOD methodology is applied to several simulation snapshots to cubic box mocks at different redshifts.

In this work, we obtain our target galaxy clustering measurements from the MXXL BGS mock. We measure the real-space correlation function from the cubic box mock made from the $z=0.2$ snapshot. This snapshot is chosen as this is the median redshift of the DESI BGS. Since we use the full cubic box, the volume is large, and we can obtain precise measurements that are not affected by cosmic variance.

\subsubsection{DESI BGS One-percent survey}
\label{sec:desi_bgs_data}
The DESI One-percent survey was conducted at the end of survey validation, and before the start of the main 5-year survey. The one-percent survey observed a footprint composed of 20 circular `rosettes', covering an area of $\sim 140~\sqdeg$ to very high completeness. The same target selection as the main survey was used, with similar exposure times \citep{DESI2023SurveyValidation}. 

Fibre assignment completeness is corrected for in the catalogue by applying a completeness weight, $w_\mathrm{comp}$. This weight is determined from 128 alternative realisations of the fibre assignment algorithm, in addition to the real survey. For each galaxy, this weight is given by
\begin{equation}
    w_\mathrm{comp} = \frac{129}{N_\mathrm{assigned} + 1},
\end{equation}
where $N_\mathrm{assigned}$ is the number of alternative realizations in which the target was assigned a fibre \citep{DESI2023DataRelease}. Note that this individual inverse probability weighting is not unbiased on very small scales. However, the one-percent survey data is highly complete, so these weights are typically close to 1. 
A pairwise inverse probability (PIP) weighting can in principle provide an unbiased correction \citep{Bianchi2017,Smith2019PIP,Bianchi2020,Mohammad2020}, and will be provided in future DESI data releases \citep{Lasker2024}.
The BGS one-percent catalogues also provide FKP weights, $w_\mathrm{FKP}$, for the complete BGS-BRIGHT sample, which reduce the variance in the correlation function measurements where there are variations in the galaxy density \citep{FKP1994}. We do not apply any FKP weighting to our BGS clustering measurements, since we use magnitude threshold galaxy samples with a constant number density. 

In Section~\ref{sec:one_percent_comparison}, we compare the clustering of our AbacusSummit mock with measurements from the One-Percent survey. In this work, we do not fit HODs directly to this dataset, since the volume is small, and there are large fluctuations in the clustering due to cosmic variance. In future work, we will adapt the HOD fitting method to apply it directly to the larger DESI Year 1 dataset. This is discussed further in Section~\ref{sec:mock_improvements}.

\subsection{Fitting method}
\label{sec:hod_fitting}

HOD fitting was performed using the \textsc{emcee} code, a Markov chain Monte Carlo sampler \citep{emcee}. All 17 HOD meta-parameters were fitted at once with target clustering and number densities taken from the BGS mock catalogue described in \citet{Smith2017} at absolute $r$-band magnitude limits in intervals of 0.5 from -18 to -22. These clustering and number density values were themselves tuned to fit the results from the Sloan Digital Sky Survey (SDSS) \citep{Abazajian2009} \& the Galaxy and Mass Assembly (GAMA) survey \citep{Liske2015} during creation of the original mock catalogue \citep{Smith2017}. The halo catalogue corresponding to the $z=0.2$ snapshot was used in all cases.

Clustering bias becomes scale-independent on large scales; this causes a problem if the shape of the correlation function of the simulation and target data have different shapes, due to the difference in cosmology. This can never be fit well and therefore throws off the full HOD parameter fit. We adjust the target data to match the shape of the simulation large-scale correlation function by using the Zel'dovich clustering prediction at scales above $8~h^{-1}\textrm{Mpc}$. 
This is evaluated from the Zel'dovich matter power spectrum using the \textsc{nbodykit} package \citep{Hand2018}.
The rescaled correlation function is given by
\begin{equation}
\xi_\mathrm{rescaled}(r) = \xi_\mathrm{MXXL}(r) \frac{\xi^\mathrm{Zel}_\mathrm{Abacus}(r)}{\xi^\mathrm{Zel}_\mathrm{MXXL}(r)} \frac{\xi^\mathrm{Zel}_\mathrm{MXXL}(r=8)}{\xi^\mathrm{Zel}_\mathrm{Abacus}(r=8)},
\end{equation}
where the superscript `Zel' indicates the Zel'dovich correlation function, and subscripts `MXXL' and `Abacus' indicate the cosmologies. On small scales $r < 8~\hMpc$, no rescaling is applied.

In addition, a volume correction was used to ensure that the target luminosity function was reproduced despite the changing size of volume elements in different cosmologies. 
The magnitude thresholds are first shifted to 
\begin{equation}
\magr^\mathrm{rescaled} = \magr + 5 \log_{10} \left( \frac{D_L^\mathrm{Abacus}}{D_L^\mathrm{MXXL}} \right),
\end{equation}
where $D_L$ is the luminosity distance to $z=0.2$, computed in the two cosmologies. The target luminosity function (which is the same luminosity function used in the construction of the MXXL mock) is then used to obtain the number density of galaxies brighter than each rescaled magnitude threshold, $\Phi(> \magr^\mathrm{rescaled})$. This number density is then rescaled by the ratio of the volume element of the two cosmologies,
\begin{equation}
\Phi^\mathrm{rescaled} = \Phi \cdot \frac{\mathrm{d}V_\mathrm{MXXL}}{\mathrm{d}V_\mathrm{Abacus}}
\end{equation}
The volume element in each cosmology is
\begin{equation}
\mathrm{d}V = \frac{c (1+z)^2 D_A^2(z)}{H(z)} \mathrm{d}\Omega \mathrm{d}z
\end{equation}

The clustering fit was limited to a maximum scale of $50~h^{-1}\textrm{Mpc}$. At scales larger than this, BAO features become important and this shape cannot be fit to by modifying HOD parameters as it is an intrinsic feature of the cosmology.

The likelihood function used in the \textsc{emcee} fitting contained a contribution from both the clustering and the number density expected from the HOD parameters,
\begin{equation}
    \mathcal{L}(M) = -\frac{1}{2}\left(\sum_{r}{\left(\frac{\xi(r,M) - \xi_\mathrm{t}(r,M)}{\sigma_{\xi}(r)}\right)^{2}} + \left(\frac{n(M) - n_\mathrm{t}(M)}{\sigma_{n}}\right)^{2}\right)
\end{equation}
where $\mathcal{L}(M)$ is the likelihood for a particular magnitude limited sample $M$. The subscript $\mathrm{t}$ is used to represent the target values for the clustering, $\xi$, and number density, $n$.
The total length of the data vector is 677, where for each of the 9 magnitude thresholds ($M_r<-22.0, -21.5, ..., -18.0$) there is a number density, and 74 correlation function bins between $0.01 < r < 50~\hMpc$. In addition, to improve the luminosity function at faint magnitudes, number density constraints are also included for $M_r<-17.5$ and $M_r<-17.0$.

The total magnitude of the likelihood function is not important for the position of the best fit HOD parameters, but it can affect how \textsc{emcee} explores the parameter space, if the likelihood function is too steep then the walkers can become stuck in local maxima and if it is too shallow then they may not converge on the best fit solution. Through testing we were able to establish values for $\sigma_{\xi}(r)$ and $\sigma_{n}$ which produced robust fits with the correct balance between fitting the clustering and the number density. $\sigma_{\xi}(r)$ and $\sigma_{n}$ were the same for all nine of the magnitude fits. 
We choose to use a constant fractional error for $\sigma_{\xi}(r)$, which is 7 \%. This value was chosen to avoid overfitting to noise at very small scales. For $\sigma_{n}$, we also use a constant fractional error of 1 \%. 
This is chosen to avoid overfitting to either the number density or the correlation function.
We did not use a full error covariance matrix when finding the best-fit HOD parameters. This is because we were not trying to establish the likelihood surface for different HODs describing the data, but instead trying to find one set of HOD parameters which produces a good fit to the desired clustering measurements and luminosity function.

\subsection{Best-fitting HODs}
\label{sec:best_fitting_hods}

\subsubsection{Best-fitting Meta-parameters}

\begin{figure*}
    \includegraphics[width=2\columnwidth]{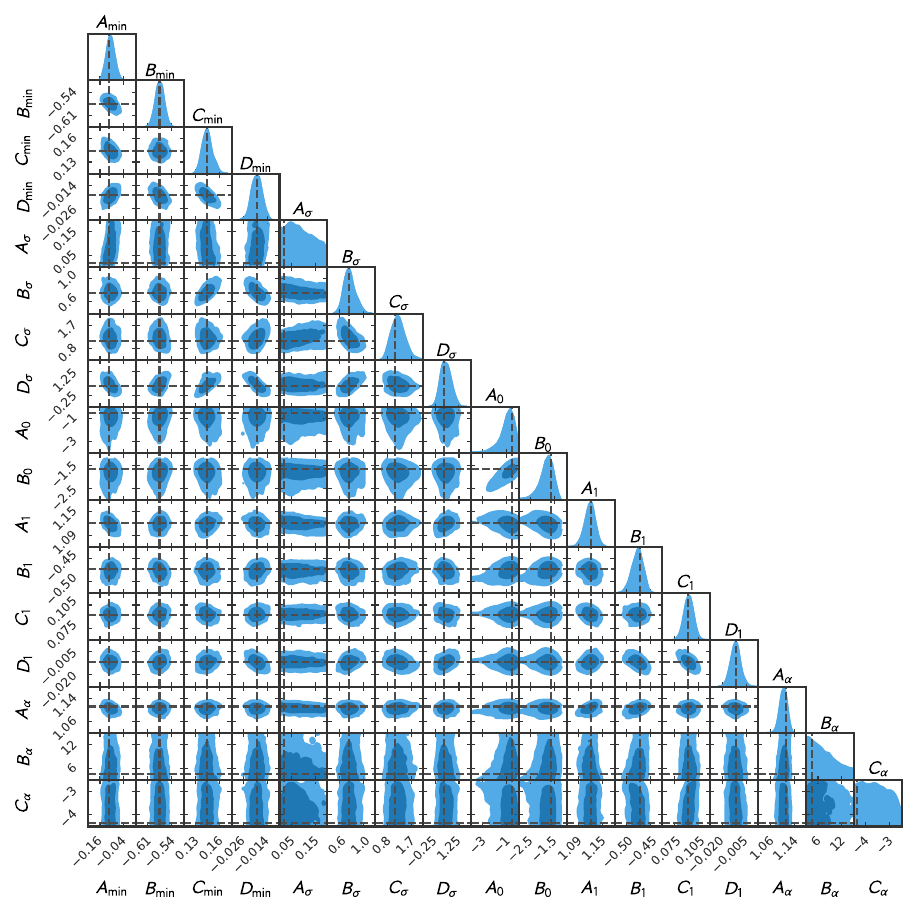}
    \caption{A corner plot from the \textsc{emcee} fitting chain of the 17 meta-parameters 
    describing the smoothly varying HOD curves. This illustrates the correlations between HOD parameters even with our assumption of uncorrelated clustering errors.
    Best-fit values are shown by the vertical and horizontal dashed lines. This example comes from the primary AbacusSummit cosmology. This figure was generated using the \textsc{pygtc} package \citep{Bocquet2016}.}
    \label{fig:corner_hod}
\end{figure*}

The output of the fitting procedure is not the HOD parameters directly, but a set of meta-parameters describing the evolution of smooth curves describing the HOD parameters as a function of absolute magnitude limit. Figure~\ref{fig:corner_hod} shows a corner plot of all 17 parameters described in Equations~\ref{eq:hod_Mmin}-\ref{eq:hod_alpha} in the case of the primary AbacusSummit Planck cosmology. The corner plot shows the parameter space explored by the \textsc{emcee} fitting chain after an appropriate amount of burn-in is discarded. The best-fit values are indicated by the dashed lines. Correlations between parameters indicate that degeneracies exist as the parameters vary along that axis. Note that since we did not use a full covariace, caution should be used when interpreting these degeneracies.

The majority of the meta-parameters are well constrained, except for the parameter $A_\sigma$, which sets the minimum value of $\sigma_{\log M}$ at faint magnitudes, and $B_\alpha$ and $C_\alpha$, which describe the exponential increase of $\alpha$ at bright magnitudes.
Reference to Equations~\ref{eq:hod_sigma}~and~\ref{eq:hod_alpha} explain the lack of constraint, because changes in these parameters have very little effect on the shape of the HOD curve in certain regimes. In addition, there is a long negative tail on $A_0$ and $B_0$. This is because $M_0$ does not affect the shape of the HOD curve if it is at sufficiently low mass that it applies when satellites have little impact.

Note that we are only fitting the HODs using diagonal errors. Our aim is to produce mocks that reproduce the galaxy clustering, to be useful for the analyses within DESI, and not to be used to determine the intrinsic errors on the HOD parameters.
The errors illustrated by the contours of Figure~\ref{fig:corner_hod} might change if a full covariance matrix is used.

\subsubsection{Best-Fitting HOD Curves}

\begin{figure}
    \includegraphics[width=\columnwidth]{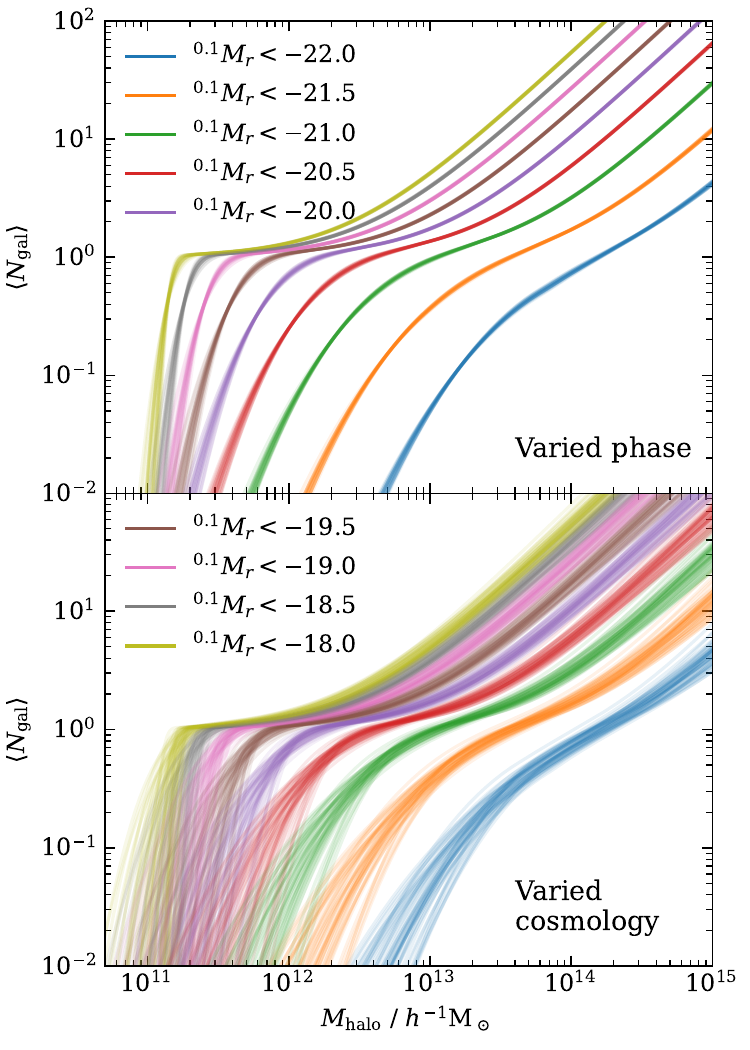}
    \caption{\textit{Top panel}: best-fitting HODs of the 25 simulations with the same c000 Planck cosmology, but with different initial conditions. Each line represents a single HOD fit. The HOD curves are highly consistent with one another for all the samples, with significant differences emerging only at the low mass tails.
    \textit{Bottom panel}: Variation of the best-fit HODs when using AbacusSummit simulations with different cosmologies. Each line represents a single HOD fit. The HOD variation is larger than the case where cosmology is held constant as in the upper panel. We only plot the HODs for AbacusSummit cosmologies in the emulator grid, to show the scatter around the primary cosmology.}
    \label{fig:best_fitting_hods}
\end{figure}

The shape of the HOD curves that are produced from the best-fit meta-parameters in the Planck cosmology case is shown in the upper panel of Figure~\ref{fig:best_fitting_hods}, for galaxy samples with different absolute magnitude thresholds. The contributions from both the central and satellite HOD components are combined into a single curve. There are no unphysical overlaps between the HOD curves as this is disallowed by the fitting procedure.
The best-fitting HODs are shown for each of the 25 Planck cosmology AbacusSummit boxes, showing that the results are very consistent, with little scatter between them.

\subsubsection{Comparison to Target Data}

\begin{figure}
    \includegraphics[width=\columnwidth]{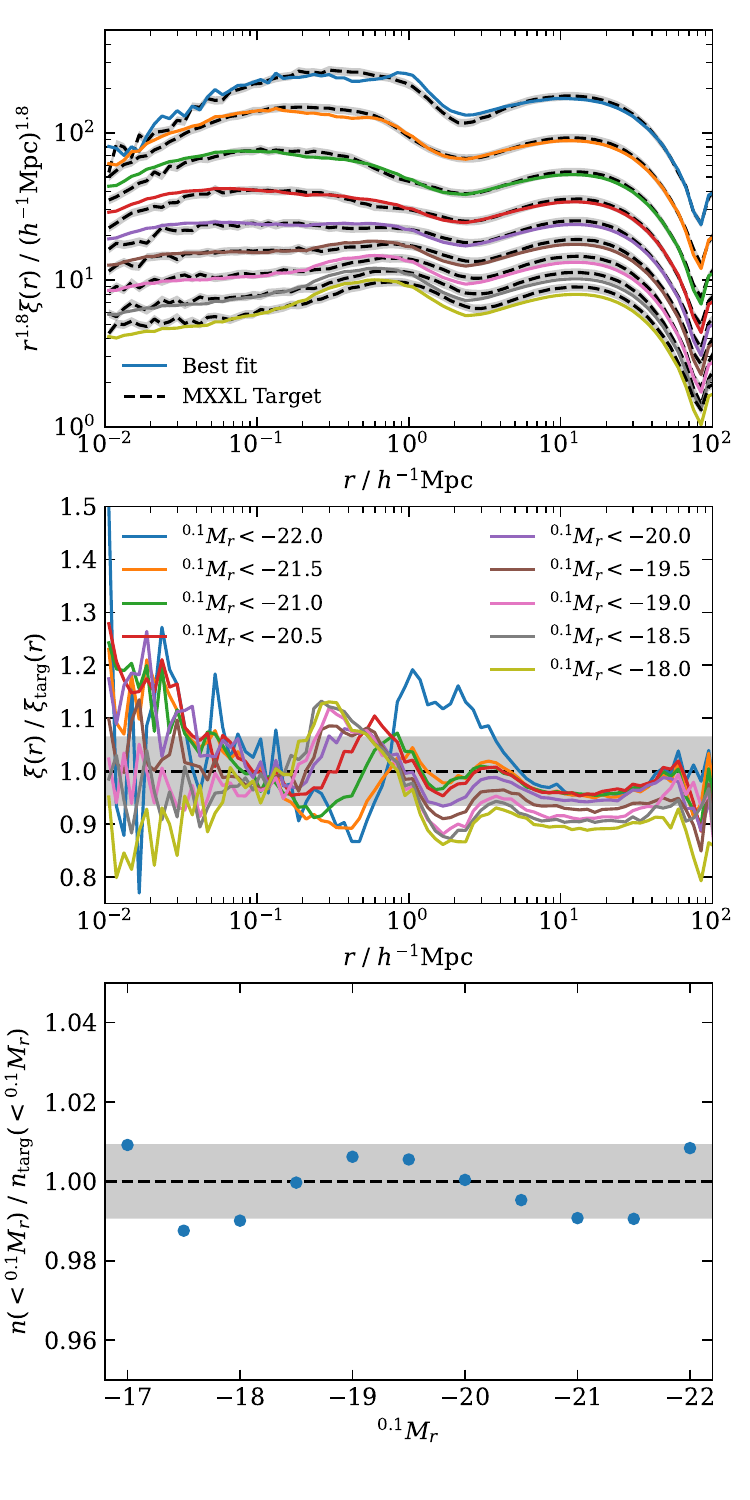}
    \caption{\textit{Top panel}: Best-fitting projected correlation function, $\xi(r)$, for the primary `c000' Planck cosmology (solid curves). Each curve is for a different magnitude threshold sample, as indicated in the legend in the middle panel. The target MXXL clustering is shown by the dashed black curves, which have been rescaled above $r=8~\hMpc$ to correct for the difference in cosmology. The black shaded region indicates the constant fractional error assumed during the HOD fitting procedure. For clarity, the curves have been offset by 0.1 dex, relative to the ${}^{0.1}M_r<-20.0$ sample.
    \textit{Middle panel}: the ratio of the best-fitting correlation functions to the target MXXL clustering measurements. 
    \textit{Bottom panel}: The ratio of the predicted and target number densities of the same magnitude threshold samples. The shaded regions shows the fractional uncertainties assumed when fitting.}
    \label{fig:target_2pcf_n}
\end{figure}

In addition to exploring the values of the best-fitting meta-parameters and shape of the HOD curves produced from them, we have also compared the quality of the best-fit with regards to the target data. Since we have fit to target data for both the clustering and the number density simultaneously, the quality of both of these aspects of the fit should be investigated.

The top panel of Figure~\ref{fig:target_2pcf_n} shows the predicted two-point clustering for the c000 Planck cosmology compared to the target measurements for the magnitude threshold samples used in the fitting procedure, where the shaded region indicated the constant fractional error that was assumed. The ratio is shown in the middle panel. The clustering measurements are mostly within this error. 
However, there is an offset on large scales, where the best-fitting clustering is systematically lower than the target clustering, by up to 10\% for the faintest magnitude threshold sample. The features above $\sim 50~\hMpc$ are outside the range of the fitting because these are caused by BAO which cannot be mitigated by varying HOD parameters. Different behaviour is seen in the residuals on small scales, depending on the magnitude limit of the sample. The brightest sample displays a peak relative to the target clustering at $\sim 2~\hMpc$, with a dip $\sim 0.4~\hMpc$, and this behaviour is inverted for the faintest samples. Below $0.1~\hMpc$, the slope of the best-fitting correlation function is different to the target, leading to differences of up to 20\% at $0.01~\hMpc$. 

In the bottom panel of Figure~\ref{fig:target_2pcf_n}, the ratio of the predicted number density to the target value is shown as a function of the magnitude limit of the sample, with the fractional error indicated by the shaded area. There is good agreement for all the samples that were included in the fitting procedure, where the number density for almost all the samples is within the error.

\subsection{Varying the cosmology}
\label{sec:varying_cosmology}

Before describing the production of the mock galaxy catalogues, we shall firstly test the robustness of the HOD fitting procedure, firstly by applying it to multiple simulations in the same cosmology, and then explore the variation of HOD parameters with cosmology.
The HOD fitting procedure described in Section~\ref{sec:hod_fitting} was applied to all base resolution AbacusSummit boxes, including all 25 c000 Planck cosmology boxes, and the boxes in other cosmologies. When applying the method to the different cosmologies, different cosmology rescaling factors must be applied to the target number densities and correlation functions.

An initial check that we performed was to find the level of sample variance that exists for HOD fits on multiple independent simulation boxes in the same cosmology. There are 25 AbacusSummit simulations in the base Planck $\Lambda$CDM cosmology (see Section~\ref{sec:abacussummit_simulations}) that were produced from initial conditions with different phase information. If there are large differences in best-fit HODs produced by the fitting procedure on these boxes, then it implies that there is a lack of robustness in the fitting procedure.

The best-fitting HODs are shown in the upper panel Figure~\ref{fig:best_fitting_hods}. There is a high degree of consistency between the HODs. Differences between HODs emerge at the low mass tails where there are very few galaxies per halo. HOD differences at these scales have a small effect on the properties of galaxy catalogues because of the low occupation. Certain points have very little scatter. These locations are where the amplitude of the HOD has the greatest effect on the number density, which is constrained in the fitting process.

Creating mocks in different cosmologies allows us to test the impact of cosmological parameters on mock observations. Cosmological parameter recovery is an important application of mock catalogues that requires mocks in different cosmologies. Exploring how the HOD parameters vary with cosmology informs us how our models of the galaxy-halo connection may change. 

The lower panel of Figure~\ref{fig:best_fitting_hods} shows the variation in HOD shapes using simulations with different cosmologies. This variation is larger than the sample variance when the same cosmology is used. The cosmology emulator grid is sufficiently wide to lead to significant differences in the best-fit HODs. This justifies our choice to refit the HODs for simulations that used different cosmologies.

\begin{figure*}
    \includegraphics[width=\columnwidth]{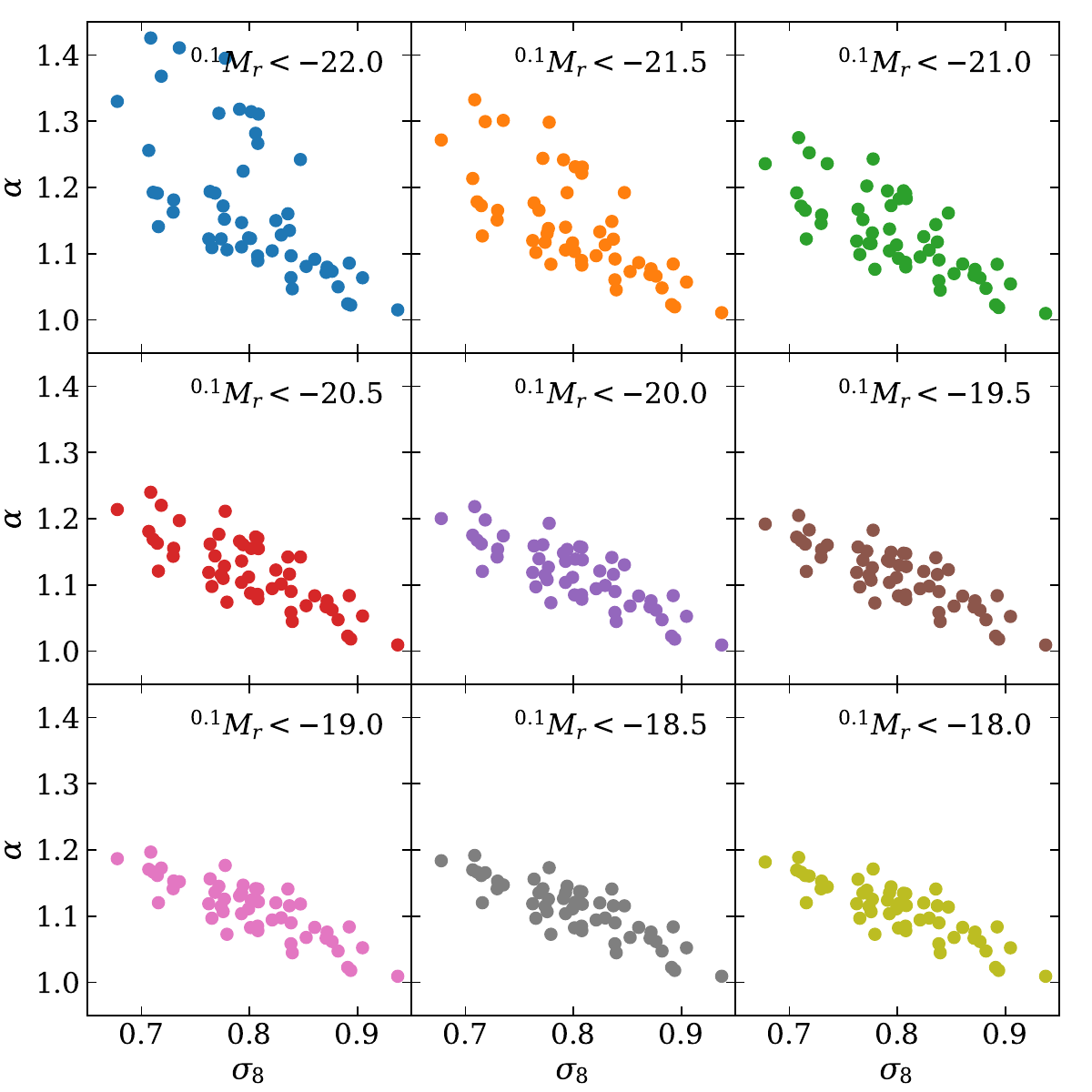}
    \includegraphics[width=\columnwidth]{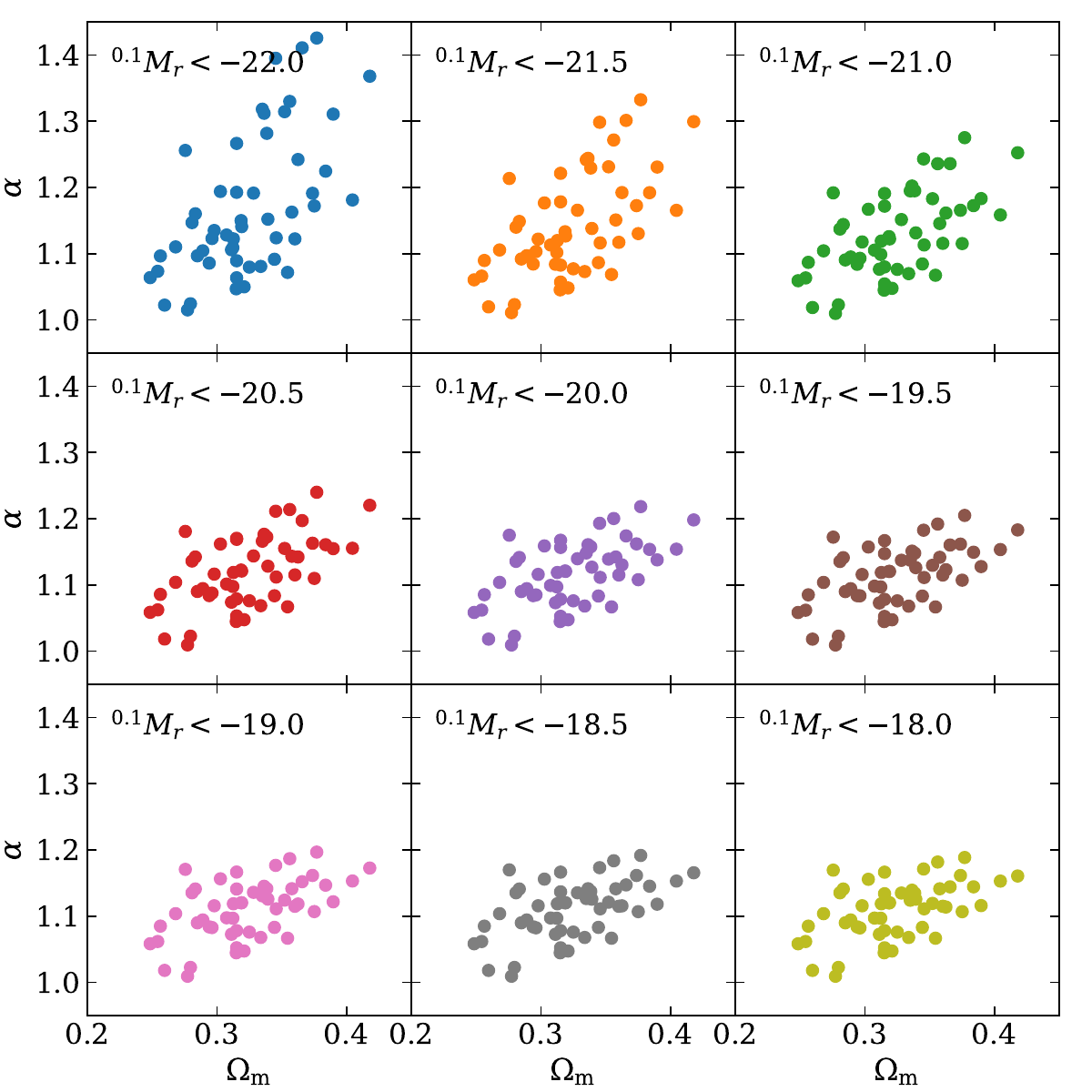}
    \caption{\textit{Left panel:} best-fitting HOD parameter $\alpha$ vs $\sigma_8$, from the AbacusSummit cosmologies in an emulator grid around the primary Planck cosmology, for different magnitude threshold samples. A negative correlation can be seen.
    \textit{Right panel:} best-fitting $\alpha$ parameter vs $\Omega_\mathrm{m}$. Here, a positive correlation is seen.}
    \label{fig:s8_alpha}
\end{figure*}

There is correlation between the best-fitting HOD parameters and the cosmological parameters of the simulation. As an example, the left panel of Figure~\ref{fig:s8_alpha} shows $\alpha$ plotted against $\sigma_8$ for different magnitude threshold samples. A clear negative correlation can be seen, where $\alpha$ is $\sim 1$ for the highest values of $\sigma_8$, but is larger for small $\sigma_8$. In addition, this correlation becomes steeper for the brightest samples, but with a larger scatter.
$\sigma_8$ sets the amplitude of the initial power spectrum at a fiducial scale of $8~\hMpc$. Simulations run with higher $\sigma_8$ will produce a halo catalogue with stronger clustering. Therefore, in order to reproduce the same target clustering, the galaxy clustering bias must be reduced in these cases. This can be achieved by reducing the number of satellite galaxies, because they are located in the highest mass (and hence most biased) halos. Therefore it makes intuitive sense that $\sigma_8$ should correlate with lower $\alpha$, if $M_0$ and $M_1$ do not change significantly. 
For the brightest samples, it is likely that the increase in the slope and scatter is due to the weak constraints on the meta-parameters $B_\alpha$ and $C_\alpha$. Our HOD model has too much freedom, since these are only constrained by the brightest few samples. In the future, we will modify the HOD model to reduce the number of meta-parameters for $\alpha$. We note that we have ignored the covariance in the correlation function estimates, but given the explanation above, we do not expect this correlation to be strongly affected.

A second example in the right panel of Figure~\ref{fig:s8_alpha}, which shows $\alpha$ plotted against the matter density $\Omega_\mathrm{m}$. Here, a positive correlation is seen, where in cosmologies with higher $\Omega_\mathrm{m}$, larger values of $\alpha$ are obtained. There are many degeneracies between the different HOD parameters, so it can be more revealing to see the effect of cosmology on the full occupation functions, rather than an individual parameter. 

\begin{figure}
    \includegraphics[width=\columnwidth]{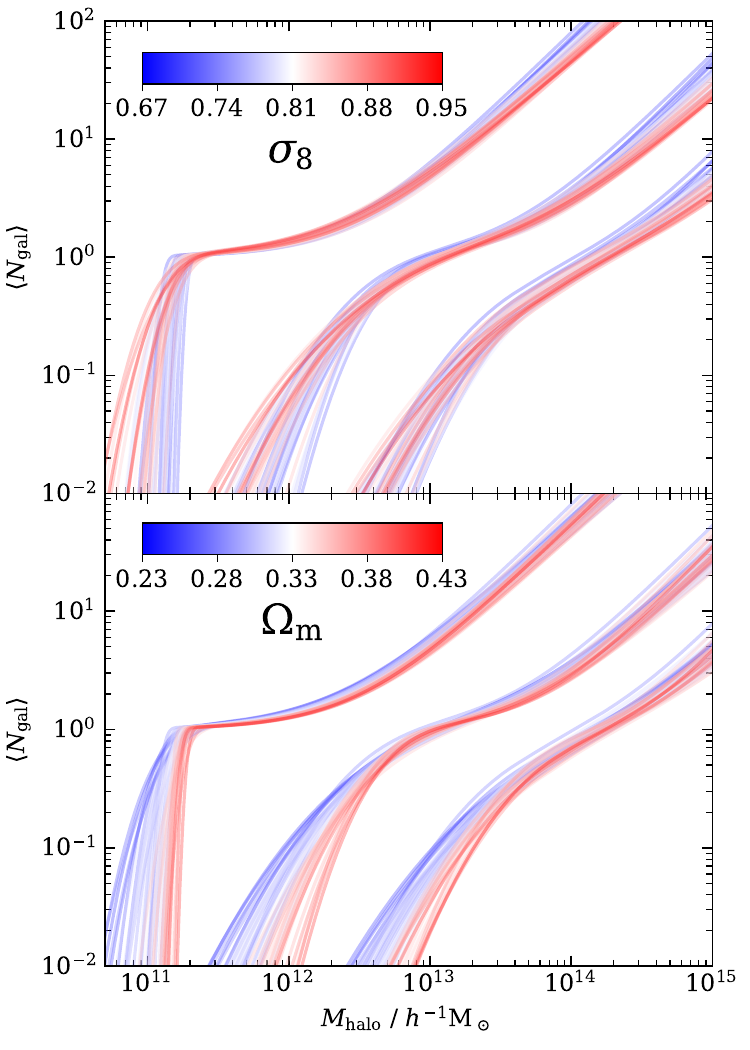}
    \caption{\textit{Top panel:} best-fitting HOD curves from the AbacusSummit simulations in different cosmologies, for the $\magr<-22$, -21 and -18 samples, from right to left. Each curve is coloured by the simulation value of $\sigma_8$, with low values in blue and high values in red.
    \textit{Bottom panel:} as above, but the curves are coloured by the value of $\Omega_\mathrm{m}$.}
    \label{fig:hod_curves_cosmology}
\end{figure}

Figure~\ref{fig:hod_curves_cosmology} shows how the shape of the best-fitting HOD curves is affected by $\sigma_8$ and $\Omega_\mathrm{m}$. The HODs are the same as in the lower panel of Figure~\ref{fig:best_fitting_hods}. We only show the $\magr<-22$, -21 and -18 samples for clarity, but all samples show the same trends. For the parameter $\sigma_8$, higher values are associated with fewer satellite galaxies, and have a broader central step function. As already discussed, halos in simulations with high $\sigma_8$ are more strongly clustered, and reducing the number of satellites and placing more centrals in lower mass halos both have the effect of reducing the clustering of the galaxy catalogue. For simulations with large values of the parameter $\Omega_\mathrm{m}$, the HODs are shifted to higher masses, with a sharper step function for the central galaxies. Even though $\alpha$, the satellite power law slope, as shown in Figure~\ref{fig:s8_alpha} is higher for large $\Omega_\mathrm{m}$, the shift in the HODs to higher masses results in a total number of satellites which is smaller. 

There are too many potential combinations of cosmological and HOD parameters to investigate all the relationships in this way. These variations in HOD parameters with cosmology may be useful if one wanted to build an emulator for HOD parameters. Such an emulator could produce an estimated best-fit HOD for any cosmology that lies within the region spanned by the AbacusSummit cosmologies.

\section{Mock creation}
\label{sec:mock_creation}

In this section we provide a brief overview of the methodology for creating mocks from the AbacusSummit simulations. Galaxies are first positioned within halos in the cubic box simulations. The nested set of HODs that we have determined are used to assign each galaxy a luminosity, as in \citet{Smith2017}. In addition, the semi-empirical method of \citet{Smith2017,Smith2022MXXL} is used to assign each galaxy a rest frame $(g-r)$ colour. 
We then process the cubic box mocks to produce `cut-sky' mocks, which convert the positions of the galaxies into sky coordinates, producing a catalogue that is more realistic and similar to what DESI observes. The mocks we create are full sky, which can then be cut to cover the DESI survey footprint. For each AbacusSummit simulation, we create one mock catalogue, using a single random realisation of the best-fitting HODs which reproduce the target clustering and luminosity function.

\subsection{Cubic Box Mock Catalogues}

Following the Monte Carlo method of \citet{Smith2017}, we use our best-fitting HODs to add galaxies to halos in the $z=0.2$ cubic box, assigning $r$-band absolute magnitudes. This is described in section~4.1 of \citet{Smith2017}. Since each HOD parameter is described as a smooth function of absolute magnitude, we can easily evaluate the HOD for any magnitude threshold, which is necessary for this method. The best-fitting HODs are also constrained to prevent any unphysical crossing of the HODs of different magnitude thresholds, which is also important when using the method to assign magnitudes. Each central galaxy is then placed at the centre of the halo, and assigned the same velocity, which is defined in AbacusSummit as the location of the centre of mass of the largest sub-halo. The number of satellite galaxies above a minimum luminosity threshold is drawn from a Poisson distribution. Magnitudes are assigned following \citet{Smith2017}, and the satellite galaxies are positioned following a NFW profile, with a random virial velocity along each direction drawn from a Gaussian with velocity dispersion $\sigma^2 = GM_\mathrm{200m} / ({2R_\mathrm{200m}})$.
We assume that the dispersion is constant, but in future, this could be modified to vary radially \citep[equation A24 of][]{Sheth2001}.

The HOD parameters $M_\textrm{min}$ and $M_1$ are modelled as cubic functions, which means that they diverge rapidly at magnitudes beyond the fitting range. To avoid this, these parameters are both extrapolated linearly (in $\log M$) beyond this range.

In addition to assigning galaxy luminosities, we also assign rest frame $(g-r)$ colours to the galaxies. This allows colour cuts to be applied to the mock and colour dependent clustering to be investigated.
We assign colours following the method of \citet{Smith2022MXXL}, which uses a parametrisation of the colour-magnitude diagram of the GAMA survey. 
This process for assigning colours is an extension of \citet{Smith2017}, which itself was based on the method of \citet{Skibba2009}. Other studies that add colours based on the prescription of \citet{Skibba2009} include \citet{Skibba2014, Carretero2015,Paranjape2015,Paranjape2021}.
It is first randomly decided whether a galaxy should lie on the red or blue sequence, which is different for central and satellite galaxies. A random colour is then drawn from a Gaussian distribution, from the fit to the GAMA colour-magnitude diagram. 
The assigned colour depends on the absolute magnitude and redshift of each galaxy, and therefore galaxies in more massive halos are more likely to be red.

The final cubic box mock is cut to an absolute magnitude limit of $\magr < -18$, which corresponds to a galaxy number density of $3.7 \times 10^{-2}~(\hMpc)^{-3}$ for the c000 Planck cosmology simulations. 

\subsection{Cut-sky mocks}
\label{sec:cut_sky_mocks}

To create a mock that is more representative of what DESI will observe, we convert the cubic box mock from the $z=0.2$ snapshot into a `cut-sky' mock. 

An observer is placed at the corner of the box, and the cubic box is replicated to cover the volume required to make a mock to a maximum redshift of $z=0.6$. Around half of the volume of the final mock fits inside a single copy of the cubic box, and it is above $z=0.37$ where the same large scale structure is repeated. Cartesian coordinates are then converted to an angular position on the sky and redshift, where the observed redshift of each galaxy includes the effect of the peculiar velocity along the line of sight. 

The cubic box we use to construct the mock is at $z=0.2$, which used HODs that were fit to reproduce the target luminosity function at the same redshift $z=0.2$. However, the BGS covers a wide redshift range of $0<z<0.6$, so we must model the evolution of the galaxy luminosity function in the cut-sky catalogue.

The target luminosity function we aim to reproduce in the mock comes from existing SDSS and GAMA survey measurements.
For $z>0.15$, the target luminosity function is a Schechter function fit to the GAMA luminosity function \citep{loveday2012}. Below $z=0.15$, the target luminosity function smoothly interpolates to that from SDSS measurements \citep{blanton2003}. The evolution of the Schechter parameters as a function of redshift is described as 
\begin{align}
M^{*}(z) &= M^{*}(z_0) - Q(z - z_0) \\
\phi^{*}(z) &= \phi^{*}(0)10^{0.4Pz},
\end{align}
where $P$ sets the evolution in number density, and $Q$ sets the magnitude evolution \citep{Lin99}. We use the values $z_0=0.1$, $P=1.8$, and $Q=0.7$, which were measured from the GAMA survey \citep{McNaught-Roberts2014}.

A rescaling is applied to the absolute magnitudes to reproduce the evolving target luminosity function, as is done in \citet{Dong-Paez2022}. In narrow redshift bins, we measure the cumulative luminosity function in the mock, and assign new magnitudes from the target cumulative luminosity function at that redshift, which correspond to the same number density for each galaxy. 
We also re-run the colour assignment algorithm on the cut-sky mock to ensure that the colour distributions also evolve smoothly with redshift, reproducing the GAMA measurements.

The full DESI BGS sample contains faint galaxies at low redshifts which live within low mass halos that fall below the resolution of the AbacusSummit simulations. We add these unresolved halos to the cut-sky mock by using the field particles (which are not in halos) as tracers. Halo masses are generated from extrapolating the measured AbacusSummit halo mass function to low masses (assuming a power law), and these halos are assigned the same position and velocity of randomly-selected field particles. The unresolved halos are assigned galaxies with magnitudes at $z=0.2$ using the same methods as the resolved halos, and the same rescaling is used to add redshift evolution.

The cut-sky mock we have created contains the rest-frame absolute magnitude, $^{0.1}M_r$ and the rest-frame ${}^{0.1}(g-r)$ colour, which we then convert to the observed quantities. 
The absolute magnitudes assigned to each galaxy are converted to an apparent magnitude following
\begin{equation}
    r = M_{r} + 5\log_{10} D_{\textrm{L}}(z_\mathrm{obs}) + 25 + k_{r}(z_\mathrm{obs}),
\end{equation}
where $D_{L}(z_\mathrm{obs})$ is the luminosity distance from the observer to the galaxy at observed redshift $z_\mathrm{obs}$, in units of $\hMpc$. $k_{r}(z_\mathrm{obs})$ is the $r$-band $k$-correction, which accounts for the shift in the band pass with redshift. We $k$-correct to a reference redshift of $z_\mathrm{ref}=0.1$. 
The distance, $D_L$, to each galaxy should only depend on its cosmological redshift, $z_\mathrm{cos}$, which does not include the effect of peculiar velocities. However, we use $z_\mathrm{obs}$ to be consistent with the data, where the cosmological redshift is not known.

The exact form of the $k$-correction depends on the filter being used and the type of object being observed. In this work, the $k$-corrections we use come from the GAMA survey and are described by fourth-order polynomials. The $k$-corrections are split into seven different rest frame $g-r$ colour bins with interpolation between the bins \citep[see section~4.3 of][]{Smith2017}. 

Similarly, galaxy colours can be transformed into the observer frame using
\begin{equation}
    g-r = {}^{0.1}(g-r) + {}^{0.1}k_{g}(z_\mathrm{obs}) - {}^{0.1}k_{r}(z_\mathrm{obs}),
\end{equation}
where $k_{g}(z_\mathrm{obs})$ is the $g$-band $k$-correction. As with the $r$-band, we use a set of colour-dependent polynomial $k$-corrections. 

Finally, an apparent magnitude cut of $r<20.2$ is applied, which encompasses both the BGS-BRIGHT and BGS-FAINT samples.

\subsection{Illustration: comparison with the DESI one-percent survey}
\label{sec:one_percent_comparison}

In this section, as an illustration, we compare number density and clustering measurements for a sample of BGS-BRIGHT ($r<19.5$) galaxies with $\magr < -20$ from the c000 Planck cosmology AbacusSummit mock with measurements from the DESI one-percent survey. In these comparisons, the same evolutionary correction has been applied to the absolute magnitudes in the data and mock, which is of the form $E(z) = Q(z-z_0)$, where $Q=0.97$ and $z_0=0.1$ \citep{McNaught-Roberts2014}.

\begin{figure}
    \includegraphics[width=\columnwidth]{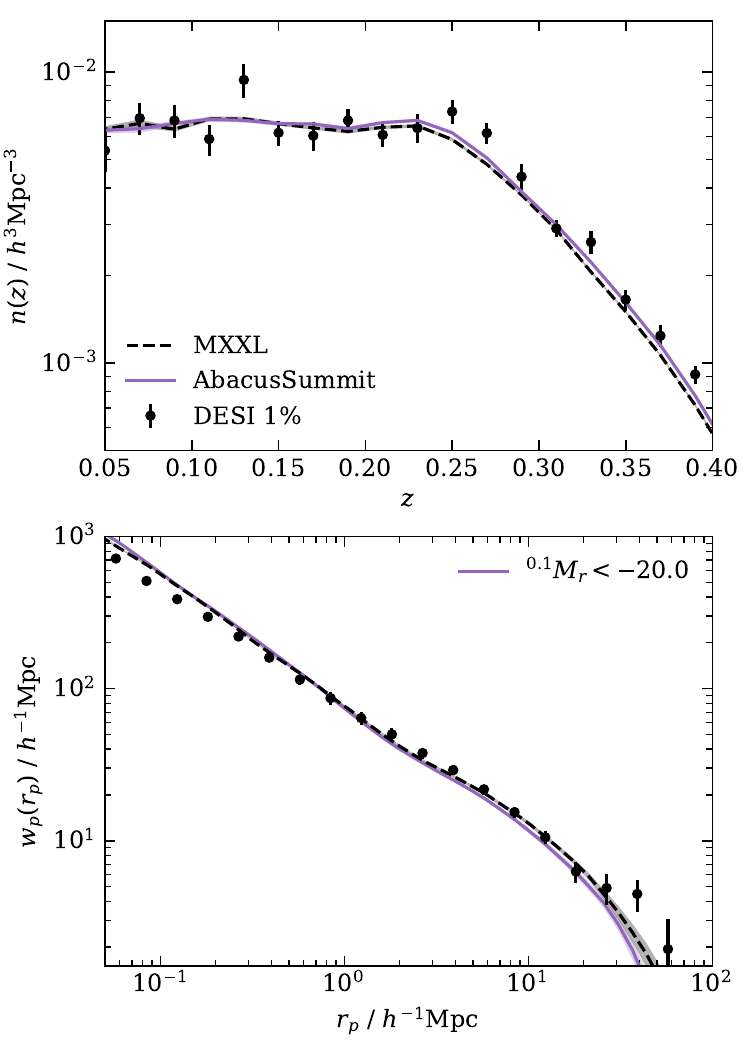}
    \caption{\textit{Upper panel}: the number density of BGS-BRIGHT ($r<19.5$) galaxies brighter than $\magr=-20.0$, as a function of redshift, for galaxies in the full-sky AbacusSummit Planck cosmology mock (purple line), the MXXL mock (dashed black line), and DESI one-percent survey measurements (points with error bars). An evolutionary correction has been applied to all absolute magnitudes, and error bars are jackknife errors with 20 jackknife regions.
    A weight is applied to the data measurements to take into account systematics and incompleteness.
    \textit{Lower panel}: projected correlation function of a volume limited sample of galaxies brighter than $\magr=-20.0$, in the redshift range $0.05 < z < 0.25$.}
    \label{fig:nz_clustering_bgs}
\end{figure}

The number density as a function of redshift is shown in the upper panel of Figure~\ref{fig:nz_clustering_bgs}, for galaxies with $\magr < -20$ in the AbacusSummit mock, MXXL mock and DESI one-percent survey. At low redshifts, since an evolutionary correction has been applied the number density is almost constant with redshift. Above $z \sim 0.25$, the density decreases since the sample become incomplete due to the $r=19.5$ cut in apparent magnitude. As expected, we see good agreement in the number densities between the AbacusSummit mock and the MXXL mock, which have the same luminosity function. A small offset is seen at high redshifts, but these mocks have different cosmologies. Both mocks agree well with the number $n(z)$ measured from the DESI one-percent survey, however the data error bars are large since the volume is much smaller than the full-sky mocks.

We measure the projected correlation function for a volume limited sample of BGS-BRIGHT galaxies with $\magr < -20$, for the same MXXL and AbacusSummit mocks and DESI one-percent survey data. The same evolutionary correction is applied, and we cut to the redshift range $0.05 < z < 0.25$, where the sample is complete (and the $n(z)$ is flat). The projected correlation function is defined as
\begin{equation}
w_p(r_p) = 2 \int_0^{\pi_\mathrm{max}} \xi(r_p, \pi) \mathrm{d}\pi,
\end{equation}
where $\xi(r_p, \pi)$ is the correlation function in bins of $r_p$ and $\pi$, perpendicular and parallel to the line of sight, respectively. Mocks and data are integrated to $\pi_\mathrm{max}=40~\hMpc$. 
The current AbacusSummit mocks are not tuned to reproduce the $w_p(r_p)$ measurements of the BGS data, so we do not expect a perfect agreement. However, this comparison demonstrates the current level of agreement, and gives insights into improvements that could be made to our HOD modelling for future mocks tuned directly to the BGS.

The projected correlation function is shown in the lower panel of Figure~\ref{fig:nz_clustering_bgs}. On intermediate scales, there is good agreement between the MXXL and AbacusSummit mocks. On large scales, there is an offset between them. The two mocks have different cosmologies, and therefore the clustering on large scales is different. This was taken into account when fitting the HODs to the real-space $\xi(r)$ measurements, but here we show the $w_p(r_p)$ measurements without any cosmology rescaling. Agreement with the one-percent survey measurements is reasonable, but not perfect. On large scales, the data is more strongly clustered than the AbacusSummit mock, and shows better agreement with MXXL. On small scales, both mocks show stronger clustering than the the one-percent survey measurements. The DESI clustering measurements are affected by fibre incompleteness on small scales, since it is not possible to place a fibre on every galaxy, particularly in dense regions such as large galaxy clusters. However, in the one-percent survey, each region is observed with multiple passes, so the completeness is very high, and a weighting is applied to correct for any small incompleteness. On these small scales, the data clustering measurements are therefore accurate, and the MXXL mock, which was tuned to SDSS and GAMA is more strongly clustered. Since AbacusSummit is fit to the MXXL clustering measurements, it also shows similar clustering to MXXL. This could potentially be improved by modifying the parameters of the NFW profiles used to position the satellites in the mock. Alternatively, fitting HODs directly to BGS clustering measurements could produce fits with fewer satellites, which would also reduce the small-scale clustering.

\subsection{Limitations and improvements for DESI}
\label{sec:mock_improvements}

We have presented a set of HOD mock catalogues for the DESI BGS, produced from the AbacusSummit simulations. The HODs have been fit to real-space $\xi(r)$ measurements from the MXXL mock, and to number densities from the SDSS and GAMA surveys. The aim of this work is a proof of concept that the fast HOD fitting method, using halo tabulation, can be used to fit HODs to multiple magnitude threshold samples simultaneously. We have validated that the best-fitting $\xi(r)$ and number densities are mostly within the errors assumed. Using these HODs to create a mock, we find good agreement in the number densities compared to the DESI one-percent survey, but there are differences in the clustering measurements which we aim to improve in future work by fitting the HODs directly to BGS data.

The DESI survey will soon be releasing data from the first full year of the survey. This large dataset will be ideal for tuning our future mocks, covering a much larger area on the sky than the one-percent survey, with smaller uncertainties in the number density and clustering measurements. However, currently there are limitations to our HOD fitting method, which we aim to address in future work.

In the current mock, we fit to real-space $\xi(r)$ measurements. In the real data, this is not available, and we will instead fit to the projected correlation function $w_p(r_p)$. To do this, the HOD fitting method must be modified to compute halo pair counts in bins of $r_p$ and $\pi$, and the correlation function $\xi(r_p,\pi)$ can be computed and integrated to obtain $w_p(r_p)$. This adds an extra dimension to the arrays of halo pair counts, slowing down each step in the MCMC chain. The effect of velocities also need to be included in the $\pi$-direction. While the increase in dimensionality reduces the speed, this can be compensated for by reducing the number of mass bins. We have checked that our fits are robust when reducing the total number of mass bins from 120 to 30.

Our HOD model assumes that the number of galaxies in each halo depends only on the halo mass. Recently, assembly bias was detected in the GAMA data in \citet{Alam2024}, and \citet{Pearl2023} detected a signal in counts-in-cylinders measurements from the BGS one-percent survey data. In future work, the HOD model could be extended to include assembly bias, enabling a cross-check of these results.

Currently when fitting the HODs, we assume a constant fraction error, both in the correlation function and number density measurements. This can be improved by using the uncertainties in the DESI measurements, which can be estimated e.g. using jackknife sub sampling. However, this does not take into account the covariances, so the fitting could be improved further by using a covariance matrix.
For the larger DESI Y1 dataset, covariance matrices will be produced using analytic and mock-based methods \citep[e.g.][]{Rashkovetskyi2023,Trusov2023}

To compute absolute magnitude from the DESI BGS data, we use colour-dependent polynomial $k$-corrections from the GAMA survey. These colour-dependent $k$-corrections are used to convert the observer frame SDSS $r$- and $g$-bands to the rest frame bands at $z_\mathrm{ref}$. We use these $k$-corrections since they can easily be applied to the mock galaxies where we only have $\gmr$ colours and do not have a full spectrum. However, these $k$-corrections are not sufficient for DESI, since there are differences between the DECam and SDSS bands, and the photometry is different in the north and south \citep{Dey2019,Zarrouk2022}. For individual DESI galaxies, the $k$-correction can be computed from the spectrum using the fastspecfit code \citep{MoustakasPrep}.\footnote{\url{https://github.com/desihub/fastspecfit}} Colour-dependent $k$-corrections can then be determined, to create a set of $k$-corrections that are appropriate for the DESI BGS survey.

The cut-sky mocks we have created are constructed from a single simulation snapshot at $z=0.2$. While we are able to reproduce the right evolving luminosity function by applying a rescaling to the magnitudes in the mock, there is no evolution in the underlying dark matter halos. In the AbacusSummit simulations, lightcone outputs were produced on the fly, which have been used to create halo lightcones \citep{Hadzhiyska2022Lightcone}. These can be used in the future to make lightcone mocks which avoid issues when using snapshots to build approximate lightcones, such as joining together the snapshots in shells \citep{Smith2022OnionShell}, or interpolating halo properties between snapshots \citep{Smith2022MXXL}.

\section{Conclusions}
\label{sec:conclusions}

For large galaxy surveys such as DESI, it is essential to use realistic mock galaxy catalogues to ensure the analyses are able to make unbiased cosmological measurements, by testing how well key statistics are recovered, and assessing systematics. A common method to create mock galaxy catalogues from large dark-matter-only simulations is to add galaxies to halos using a HOD model. 

The DESI BGS is a survey of low redshift galaxies with median redshift $z \sim 0.2$, consisting of the flux-limited BGS-BRIGHT sample, with $r$-band apparent magnitude limit $r=19.5$. The secondary BGS-FAINT sample extends this limit to $r=20.175$, with additional colour and fibre magnitude cuts to ensure a high redshift success rate. As was previously done for the MXXL simulation, mock galaxy catalogues with $r$-band magnitudes can be created using a set of `nested' HODs for different absolute magnitude thresholds. However, when fitting the HODs of sample independently to data clustering measurements, it is possible for the HODs of different magnitude thresholds to cross unphysically.

We have developed a fast HOD fitting method to simultaneously fit HODs for multiple magnitude threshold samples. By fitting all the samples at once, we constrain the HODs to prevent unphysical crossing of the HODs.. Halo pair counts are first tabulated as a function of halo mass, which only needs to be done once. The correlation function for a given HOD model can then be evaluated quickly from a weighted sum of the halo pair counts. Each HOD parameter is defined as a smooth function of absolute magnitude, and we fit the meta-parameters defining these smooth curves. When fitting, we include constraints on both the galaxy clustering and number densities for each of the magnitude threshold samples. 

As a proof of concept, we apply the HOD fitting procedure to the AbacusSummit simulations, using the snapshot at $z=0.2$, and fitting to real-space correlation functions, $\xi(r)$ from the previous MXXL mock, and number densities from the SDSS and GAMA surveys. Since the MXXL mock is in a WMAP1 cosmology, which is different to the AbacusSummit simulation cosmologies, we apply a rescaling to the clustering measurements on scales $r > 8~\hMpc$ so that the large-scale clustering of the two simulations matches. A cosmology rescaling is also applied to the target number densities. We first apply the HOD fitting to the 25 AbacusSummit simulations in the primary Planck cosmology. We find that there is very little scatter between the best-fitting HODs of the 25 simulations, which verifies the robustness of the fitting procedure. Most of the meta-parameters are well constrained, and the parameters with the weakest constraints only have little effect on the shape of the HODs. The number densities achieved are within the assumed errors, and the clustering measurements are mostly within the errors, although the clustering is low on large scales for the faintest samples. 

We also apply the HOD fitting procedure to the AbacusSummit simulations in a range of different cosmologies. Varying the cosmology produces much more variation in the HODs, and there are trends with the different cosmological parameters. For example, increasing the parameter $\sigma_8$ increases the amplitude of the initial power spectrum and hence the halo catalogues in these cosmologies are more strongly clustered. To match the same target galaxy correlation function, the best-fitting HODs have fewer satellites, and a broader central step function. For the parameter $\Omega_\mathrm{m}$, the best-fitting HODs in cosmologies with high $\Omega_\mathrm{m}$ are shifted to higher masses, and have a sharper central step.

We use the best-fitting HODs to create AbacusSummit mock catalogues for the DESI BGS. We first populate the cubic box at $z=0.2$ with galaxies, using the nested set of HODs to assign absolute $r$-band magnitudes following the same method used to create the MXXL mock. A Monte Carlo method is also used to assign $\gmr$ colours. The snapshot is then converted to a `cut-sky' mock by replicating the box, and converting the galaxy coordinates to a sky coordinate and redshift, where the redshift takes into account the velocity of the galaxy along the observer's line of sight. Field particles are used as tracers of halos which fall below the mass resolution of the simulation. A rescaling is applied to the magnitudes to achieve a smoothly evolving luminosity function, and colours are re-assigned so that the colour distributions also evolve smoothly. The $r$-band apparent magnitude and observer frame $g-r$ colours are calculated using colour-dependent $k$-corrections from the GAMA survey. 

As an illustration, we compare the number density and clustering of one of the base Planck cosmology cut-sky mocks with BGS measurements from the one-percent survey, for a sample of galaxies with magnitude threshold $\magr<-20$. The $n(z)$ shows good agreement, although the uncertainties in the BGS measurements are large, due to the small volume of the one-percent survey. The projected correlation function, $w_p(r_p)$ measurement in the mock is reasonable, but the data is more strongly clustered than the mock on large scales, while the opposite is true on small scales. In the future, we aim to improve these mocks by extending the tabulation method to fit directly to $w_p(r_p)$ measurements from the DESI year 1 dataset, which covers a much larger area than the one-percent survey. We also aim for future mocks to improve the errors and take into account covariances, and to use DESI $k$-corrections.

The AbacusSummit mocks we have presented here are being used within the DESI collaboration as part of the first generation of BGS mocks, which we make publicly available.


\section*{Acknowledgements}

AS, CG, SC and PN acknowledge the support of STFC grants ST/T000244/1 and ST/X001075/1.
Significant work presented in this paper is part of Cameron Grove's PhD thesis \textit{`Gravity to Galaxies -- N-body Simulations for the DESI Survey'} \citep{CameronThesis}.

This material is based upon work supported by the U.S. Department of Energy (DOE), Office of Science, Office of High-Energy Physics, under Contract No. DE–AC02–05CH11231, and by the National Energy Research Scientific Computing Center, a DOE Office of Science User Facility under the same contract. Additional support for DESI was provided by the U.S. National Science Foundation (NSF), Division of Astronomical Sciences under Contract No. AST-0950945 to the NSF’s National Optical-Infrared Astronomy Research Laboratory; the Science and Technology Facilities Council of the United Kingdom; the Gordon and Betty Moore Foundation; the Heising-Simons Foundation; the French Alternative Energies and Atomic Energy Commission (CEA); the National Council of Science and Technology of Mexico (CONACYT); the Ministry of Science and Innovation of Spain (MICINN), and by the DESI Member Institutions: \url{https://www.desi.lbl.gov/collaborating-institutions}. Any opinions, findings, and conclusions or recommendations expressed in this material are those of the author(s) and do not necessarily reflect the views of the U. S. National Science Foundation, the U. S. Department of Energy, or any of the listed funding agencies.

The authors are honored to be permitted to conduct scientific research on Iolkam Du’ag (Kitt Peak), a mountain with particular significance to the Tohono O’odham Nation.

\section*{Data Availability}

The AbacusSummit cubic box and cut-sky mocks produced in this work will be made available at \url{https://icc.dur.ac.uk/data/}.

The data points and Python scripts for reproducing the figures in this publication will be made available on Zenodo.



\bibliographystyle{mnras}
\bibliography{abacussummit_mocks} 






\bsp	
\label{lastpage}
\end{document}